\documentclass[%
reprint,
 amsmath,amssymb,
 aps,
showpacs,showkeys,]{revtex4-1}

\usepackage{graphicx}% Include figure files
\usepackage{dcolumn}% Align table columns on decimal point
\usepackage{bm}% bold math
\begin{document}

\title{Approximating the $S$ matrix for solving the Marchenko equation: the case 
	of channels with different thresholds
}

\author{N. A. Khokhlov}
\email{nikolakhokhlov@yandex.ru; nakhokhlov@gmail.com}
\affiliation{%
Peter the Great St. Petersburg Polytechnic University, St. Petersburg, Russia.
}%
\date{\today}% 

\begin{abstract}
This work extends previous results on the inverse scattering problem within the framework of Marchenko theory (fixed-$l$ inversion).
In particular, I approximate an $n$-channel $S$-matrix as a function of the first-channel momentum $q$ by a sum of a rational term and a truncated sinc series for each matrix element.
Relativistic kinematics are taken into account  through  the correct momentum-energy relation, and the necessary minor generalization of Marchenko theory is given.
 For energies where only a subset of scattering channels is open, the analytic structure of the $S$-matrix is analyzed. I demonstrate that the submatrix corresponding to closed channels, particularly near their thresholds, can be reconstructed from the experimentally accessible submatrix of open channels.The convergence of the proposed method is verified by applying it to data generated from a direct solution of the scattering problem for a known potential, and comparing the reconstructed potential with the original one. Finally, the method is applied to the analysis of $S_{31}$ $\pi N$ scattering data.
\end{abstract}
\pacs{24.10.Ht, 13.75.Cs, 13.75.Gx}
\keywords{quantum scattering, Marchenko theory, inverse problem, algebraic method, numerical solution, $\pi N$ scattering}

\maketitle

\section{\label{sec:intro}INTRODUCTION}

A major challenge in theoretical nuclear physics is the inversion of scattering data to obtain the interparticle interaction.
Existing approaches to this inverse problem (IP) can be divided into two categories. The first involves fitting the parameters of phenomenological or theoretical potentials. These methods typically constitute a nonlinear parameter-fitting procedure and are feasible only when the amount of experimental data is limited. The second, more systematic category comprises (for fixed-l partial wave analysis (PWA)) methods based on the theories of Marchenko, and of Gelfand, Levitan, and Krein \cite{Gelfand, Agranovich1963, Marchenko1977, Blazek1966, Krein1955, Levitan1984, Newton, Chadan,Kukulin2004,Mack2012}. 
These methods provide a unique local potential but require complete knowledge of the problem spectrum: the $S$-matrix from zero to infinity energy, binding energies, and normalization constants). 
 An extension of their applicability beyond nonrelativistic quantum mechanics
is achieved by expressing relativistic two-particle potential models in a nonrelativistic form \cite{Keister1991}.

The Marchenko theory  employed here has been previously generalized to cases of coupled  partial waves with different thresholds \cite{Cox1975a,Cox1975b}. 
For completeness, I mention Ref.~\cite{Braun2003}, which addresses a one-dimensional multichannel IP. While not directly related to this work, it shares certain conceptual similarities.
However, for our case, explicit potential calculations from PWA data have only been carried out for equal thresholds \cite{Geramb1994, Kohlhoff1994, Khokhlov2006, Khokhlov2007}. 
Implementations of the Marchenko theory have relied on approximating the partial $S$-matrices using rational-fraction expansions. This approximation allows the input kernel of the Marchenko equation to be expressed as a finite separable series. In this case, the Marchenko equation admits an analytic solution, resulting in potentials—Bargmann-type potentials—that can be written explicitly in terms of Riccati-Hankel functions.
A sufficiently accurate approximation of the $S$-matrix via a rational-fraction expansion presents a challenge. The primary difficulty is the emergence of spurious poles in the $S$-matrix near or on the physical momentum $q$-axis. These unphysical poles inevitably arise when fitting high-precision PWA data, as in $NN$ scattering analyses up to 3 GeV. To circumvent this issue, the PWA data must be artificially smoothed, a procedure that inevitably degrades the accuracy of the subsequent IP solution.

Two numerical methods have recently been proposed for solving the single-channel Marchenko equation: one based on decomposing the input kernel into a separable series of isosceles triangular-pulse functions \cite{MyAlg2,MyAlg3}, and another based on approximating the $S$-matrix by a sum of a rational term and a truncated sinc series \cite{MyAlg4}.
Both methods are capable of reproducing arbitrary $S$-matrices without introducing spurious poles and are convergent.

This work presents two key developments: a generalization of the method in \cite{MyAlg4} to the case of coupled channels with different thresholds, and an extension of Marchenko theory that accounts for relativistic kinematics.
The multichannel $S$-matrix behavior  in the inter-threshold region is analyzed in detail. This analysis clarifies the analytic structure of the $S$ matrix near thresholds and demonstrates the possibility of continuing the $S$-matrix from the experimentally accessible region above threshold to the region below.

\section{Relativistic corrections}
The developed inversion method is applied to analyze $\pi N$ scattering data at energies where relativistic effects are significant. These effects are incorporated within the framework of relativistic quantum mechanics for systems with a fixed number of particles \cite{Keister1991}. The analysis assumes coupling only between two-particle channels. In such a system, the wavefunction is an eigenfunction of the mass operator and can be represented as a product of external and internal components \cite{Lev,Khokhlov}. The internal WF, denoted $\chi$, is itself an eigenfunction of the mass operator and satisfies the following equation:
\begin{equation}
	\label{rel00} \left[ {\hat M^{(0)}}+{\hat V}_{int} \right]\chi = M\chi,
\end{equation}
where ${\hat M_{0}}$ is the diagonal nonineracting mass operator 
\begin{multline}
	%\begin{equation}
	\label{M000}
	{\hat M^{(0)}} =
	\begin{bmatrix}
		{\hat M}_{1} & 0 & \cdots & 0 \\
		0 & 	{\hat M}_{2} & \ddots & \vdots \\
		\vdots & \ddots & \ddots & 0 \\
		0 & \cdots & 0 & 	{\hat M}_{n}
	\end{bmatrix},\\
	{\hat M}_{\alpha}=w_{\alpha 1}(q)+w_{\alpha 2}(q)\\
	\equiv\sqrt{\hat{q}^2+m_{\alpha 1}^2}+\sqrt{\hat{q}^2+{m}_{\alpha 2}^2};
	%	\begin{bmatrix}\sqrt{\hat{q}^2+m_{N}^2}+\sqrt{\hat{q}^2+m_{1}^2} & 0 \\ 
		%	0 & \sqrt{\hat{q}^2+m_{N}^2}+\sqrt{\hat{q}^2+{m}_{2}^2}\end{bmatrix},
	%	{\hat M_{0}}+V_{int} \right]\chi = M\chi,
\end{multline}
%\end{equation}
and ${\hat V}_{\text{int}}$ is the matrix interaction operator acting only through
internal variables (spins and relative momentum); $\hat{q}$ is the
momentum operator of one of the particles in the center-of-mass
frame (relative momentum). $M$ is the mass eigenvalue of the system, $m_{\alpha k}$  is the  mass of the $k$th particle in the $\alpha$th channel. 
The thresholds $W_{\alpha}= m_{\alpha 1}+m_{\alpha 2}$, and 
   $W_{1}  < W_{2} < \dots <W_{n} $. 
The momentum $q_{1}$ in the first channel corresponding to the $\alpha$th threshold is determined from
$W_{\alpha}=\sqrt{{q_{1}}^2+m_{1 1}^2}+\sqrt{{q_{1}}^2+{m}_{1 2}^2}$. 

Rearrangement of Eq.~(\ref{rel00}) gives
\begin{equation}
	\left[ {\hat{q}^2 + {\hat V}} \right]\chi = Q^2\chi,\label{eq_Shred}
\end{equation}
where $Q$ is the diagonal matrix of channel momenta 
\begin{equation}
	\label{Q}
	{Q} =
	\begin{bmatrix}
		{q}_{1} & 0 & \cdots & 0 \\
		0 & 	q_{2} & \ddots & \vdots \\
		\vdots & \ddots & \ddots & 0 \\
		0 & \cdots & 0 & 	{q}_{n}
	\end{bmatrix}
	%\begin{bmatrix}q_{1} & 0 \\ 
	%	0 & q_{2} \end{bmatrix},
%	{\hat M_{0}}+V_{int} \right]\chi = M\chi,
\end{equation}
with
\begin{equation}
\label{rel3} q_{\alpha}^2=\frac{M^2}{4}-\frac{m_{\alpha 1}^2+m_{\alpha 2}^2}{2}+\frac{(m_{\alpha 1}^2-m_{\alpha 2}^2)^2}{4M^2}.
\end{equation}
%
%$i=1,2$ correspond to $\pi$ and $\eta$.
${\hat V}$ is an operator acting like ${\hat V}_{int}$ only through internal variables. The simplest case of Eq.~(\ref{rel3}) is  $m_{\alpha 1}=m_{\alpha 2}=m_{\alpha}$ then
\begin{equation}
\label{simple_rel3} 
q_{\alpha}^2=\frac{M^2}{4}-m_{\alpha }^2\equiv 	q_{1}^2-\Delta_{\alpha}
\end{equation}
corresponds to the non-relativistic  approximation considered in \cite{Newton,Taylor}.

The relationship between ${\hat V}$ and the original interaction operator ${\hat V}_{\text{int}}$, as well as that between ${\hat V}$ and the underlying fundamental interactions is nontrivial. Nevertheless, ${\hat V}$ correctly provides the nonrelativistic limit of both.

Eq.~(\ref{eq_Shred}) is formally identical to the Schr\^{o}dinger equation.This correspondence is made explicit in the quasicoordinate representation, which adopts the realization ${\hat q} = - i\, \partial / \partial {\bf r}$, and ${\hat V} = V({\bf r})$. This formal equivalence permits the application of the developed inversion algorithm, provided the requisite modifications to the Marchenko theory are made.

\section{Multichannel Marchenko inversion}
The radial multichannel Schrödinger equation,
\begin{equation}
	\label{system2by2}
\left(- \frac{d^{2}}{d r^{2}} + L(L+1)r^{-2}+V\right) \Psi(r)
=Q^{2} \Psi(r).
\end{equation}
Here, $L$ is a diagonal matrix with elements $L_{\alpha \alpha} = l_{\alpha}$ (the orbital angular momentum for $\alpha$th channel); $V(r)$ is the potential symmetric matrix; the wave function $\Psi(r)$ is the square $n\times n$ matrix \cite{Chadan,Newton,Taylor}.

The asymptotics of some regular solution ($\Psi (0)=0$) defines the $S$ matrix
\begin{equation}
	\label{asymptotics_of_sol}
	\Psi (r) \xrightarrow[r \to \infty]{}
 \frac{\imath }{2}
	\left(H^{(-)}(Qr)  - H^{(+)}(Qr)Q^{-1/2}S Q^{1/2} \right),
\end{equation}
where $H^{(\pm)}(Qr)$ are diagonal matrices, with $H^{(\pm)}_{\alpha \alpha}(Qr) = h^{\pm}_{l_{\alpha}}(q_{\alpha}r)$, and $h_{l}^{\pm}(z)$ are the Riccati-Hankel functions defined as in \cite{Taylor} (table 11.1). In nonrelativistic theory time-reversal invariance implies the symmetry of the $S$ matrix and of the potential matrix $V$ \cite{Taylor,Newton}. I assume that these properties hold under the minimal relativistic corrections of the previous section.

The Marchenko integral equation  \cite{Agranovich1963, Marchenko1977} is formally the same
\begin{equation}
	\label{f3}	F(x, y)+L(x, y)+\int_{x}^{+\infty} L(x, t) F(t, y) d t=0,
\end{equation}
but input kernel $F(x, y)$  and output kernel $L(x, y)$  are matrix functions.
The input kernel is defined as \cite{Cox1975a,Cox1975b,Chadan}
\begin{eqnarray}
	F(x, y)\nonumber \\ 
	=\frac{1}{2 \pi} \int_{-\infty}^{+\infty} 
	  dq_{1}
	H^{(+)}(Qx) 
	\rho	[1-S(q)] \rho
	H^{(+)}(Qy)  \nonumber  \\	
	+\sum_{j=1}^{n_{b}} 	H^{(+)}(\tilde{Q}_{j} x) A_{j} 	H^{(+)}(\tilde{Q}_{j} y )
	\label{f4first} 
%	\\	=\frac{1}{2 \pi} \int_{-\infty}^{+\infty}   dq_{1} H^{(+)}(Qx) Y(q) H^{(+)}(Qy). 
%	\label{f4}	
\end{eqnarray}
$\rho$ is a diagonal matrix with elements
\begin{equation}
	\label{rho_my}	
		\rho_{\alpha \alpha} =\left(\frac{q_{1}}{q_{\alpha} }
		\frac{w_{\alpha 1}(q_{\alpha})w_{\alpha 2}(q_{\alpha})
		}{ 	w_{1 1}(q_{1})w_{1 2}(q_{1})}
		\right)^{1/2}\equiv \left(\frac{dq_{\alpha}}{dq_{1}} \right)^{1/2},
\end{equation}
$\tilde{Q}_{j}$ is a matrix of purely imaginary values, and  $A_{j}$ is a symmetric matrix; both  correspond to an ordinary $j$th bound state. I assume no continuum bound states. Eq.~(\ref{rho_my}) follows from $dM=Mq_{\alpha} dq_{\alpha}/w_{\alpha 1}(q_{\alpha})w_{\alpha 2}(q_{\alpha})$ and
is a minor generalization of the results presented by J.~R.~Cox \cite{Cox1975a,Cox1975b} (for Eq.~(\ref{simple_rel3}) where $dq_{\alpha}/dq_{1}\equiv q_{1}/q_{\alpha}$). 
%, and from corresponding completeness relation for the regular solutions of Eq.~(\ref{eq_Shred}). 

The potential matrix function of Eq.~(\ref{eq_Shred}) is obtained from the output kernel 
\begin{equation}
	V(r)=-2 \frac{d L(r, r)}{d r}. \label{Vpot6}
\end{equation}
Eqs.~(\ref{system2by2})-(\ref{Vpot6}) are valid for non-singular potentials of finite range.

\section{Parameters of the $S$-matrix for coupled channels with different thresholds}
Only the open-channel $S_{+}$ submatrix ($n_{+}\times n_{+}$) of the $S$ matrix ($n\times n$), along with certain bound-state parameters ($\tilde{Q}_{j}$ and possibly $A_{j}$), 
 are directly accessible to experiment. 
The limited number of required parameters must accurately incorporate all essential properties of the $S$-matrix. Since the present inversion method relies exclusively on the $S$-matrix values for real $q_{1}$, we restrict our analysis to the corresponding properties on the physical axis.  The generalization  of case~(\ref{simple_rel3}) \cite{Taylor,Newton} to case~(\ref{rel3}) is straightforward and relies on the standard properties of the multichannel $S$-matrix:    

1. The $S$ matrix is symmetric, $S(q) = S^{T}(q)$.

2. The $S_{+}$ matrix is unitary, $S_{+}^{-1}=S_{+}^{+}$.

3. For real $q_{1}$, $S(-q_{1})=S(q_{1})^{*}$, where  $^{*}$ denotes complex conjugation. 

4. For $|M| \leq W_{\alpha}$, $q_{\alpha}=\imath |q_{\alpha}|$, $\text{Im } q_{\alpha}\geq 0$ (the physical sheet of $M$ corresponds to all  $q_{1},q_{2},\dots,q_{n}$ in their respective  upper half planes). 
For $|M| \geq W_{\alpha}$, $q_{\alpha}=sign(q_{1}) |q_{\alpha}|$. 
A detailed explanation of this analytic structure is given in \cite{Taylor} (Ch.~20-a).

To proceed further I use the following equations \cite{Calo}, which are preserved in case of Eq.~(\ref{rel3}).
For potential matrix
\begin{equation}
	V_{{\tilde r}}(r)=
	\begin{cases}
		V(r), & r\leq {\tilde r}\\
		0	 & r> {\tilde r},
	\end{cases}
	\label{V_varied}
\end{equation}
the symmetric reaction matrix $T({\tilde r})$ is defined by 
\begin{eqnarray}
	\label{Tmatrix}	
T({\tilde r})=\imath (1-S({\tilde r}))(1+S({\tilde r}))^{-1},\\
S({\tilde r})= (1-\imath T({\tilde r}))^{-1}(1+\imath T({\tilde r})),
\end{eqnarray}
where $S({\tilde r})$ is defined as $S$ in Eq.~(\ref{asymptotics_of_sol}) but for $V_{{\tilde r}}(r)$ instead of $V(r)$.
Matrix-function $T( r)$ satisfies the differential equations
\begin{multline}
	\label{Tmatrix_De}	
\frac{d}{dr}T( r)= -2 \left( J(Qr)-T(r)N(Qr)\right)Q^{-1/2} V(r)\\
\cdot Q^{-1/2} \left( J(Qr)-N(Qr)T(r)\right),
\end{multline}
%\begin{multline}
%	\frac{d}{dr}S( r)= \imath \left( S(r)  H^{(+)}(Qr)- H^{(-)}(Qr)\right)Q^{-1/2}\\
%	 \cdot  V(r)  Q^{-1/2} \left( H^{(+)}(Qr)S(r)- H^{(-)}(Qr)\right);
%\end{multline}
where 
\begin{multline}
	J(Qr)=(H^{(+)}(Qr)-H^{(-)}(Qr))/2\imath,\\
N(Qr)=-(H^{(+)}(Qr)+H^{(-)}(Qr))/2.
\end{multline}
 The boundary conditions are
\begin{eqnarray}
	\label{bounds_SandTmatrix1}	
	T(0)=0, \\ 
	T(\infty)=T=\imath (1-S)(1+S)^{-1};\\
	S(0)=1,\\ 
	S(\infty)=S=(1-\imath T )^{-1}(1+\imath T).
	\label{SfromT_inf}
\end{eqnarray}
For $n_{+}$ open channels, and omitting the obvious dependencies on $r$. I introduce diagonal block matrices 
\begin{eqnarray}
	\label{Q12_1}
	{Q} =
	\begin{bmatrix}
		{Q}_{+} &  0 \\
			0 & \imath {Q}_{-}
	\end{bmatrix},\\
		\label{Q12_J}
			J(Qr)=
	\begin{bmatrix}
		J_{+} &  0 \\
		0 &  \imath^{L+1} J_{-}
	\end{bmatrix},\\
		\label{Q12_N}
		N(Qr)=
	\begin{bmatrix}
		N_{+} &  0 \\
		0 &  \imath^{L} N_{-}
	\end{bmatrix},
	\\
		\label{Q12_H}
	H^{(\pm)}(Qr)=
		\begin{bmatrix}
		H^{(\pm)}_{c+} &  0 \\
		0 &  \imath^{L} H^{(\pm)}_{-}
	\end{bmatrix},
	\end{eqnarray}
	where $Q_{+}$, $J_{+}$, $H^{(\pm)}_{-}$,  and $N_{+}$ are real diagonal matrices $n_{+}\times n_{+}$; $Q_{-}$, $J_{-}$,  and $N_{-}$ are real diagonal matrices $(n-n_{+})\times (n-n_{+})$.
The complex factor $\imath$ is factored out where possible. 
I introduce block symmetric matrices 
\begin{eqnarray}
		V(r)=
	\begin{bmatrix}
		V_{11}(r) &  V_{12}(r) \\
		V_{21}(r) &  V_{22}(r)
	\end{bmatrix},\\
		T(r)=
	\begin{bmatrix}
		T_{11}(r) &  T_{12}(r) \\
		T_{21}(r) &  T_{22}(r)
	\end{bmatrix},
	%\\
	%	S(r)=
	%\begin{bmatrix}
	%	S_{11}(r) &  S_{12}(r) \\
	%	S_{21}(r) &  S_{22}(r)
	%\end{bmatrix}.
\end{eqnarray}
where $V_{11}$ and $T_{11}$ are matrices $n_{+}\times n_{+}$,
$V_{12}$ and $T_{12}$ are matrices $(n-n_{+})\times n_{+}$,
$V_{21}=V_{21}^{T}$, $T_{21}=T_{21}^{T}$; 
and  $V_{11}$ and $T_{11}$ are matrices $(n-n_{+})\times (n-n_{+})$.

Inserting Eqs.~(\ref{bounds_SandTmatrix1}), (\ref{Q12_1})--(\ref{Q12_N}) in
Eq.~(\ref{Tmatrix_De}) I find
\begin{multline}
	\label{dT0}
	T(d r) =d r \frac{dT( 0)}{dr}\\
	= d r	\begin{bmatrix}
		 \tilde{ T}_{0+}(dr) &  	\imath^{L+1/2}  \tilde{ T}_{0+-}(dr) \\
		\imath^{L+1/2}  \tilde{ T}_{0-+}(dr) &  	\imath T_{0-}(dr)
	\end{bmatrix},
\end{multline}
where submatrices $	 \tilde{ T}_{0+}$, $	 \tilde{ T}_{0+-}$, and $	 \tilde{ T}_{0-}$ are real. 
Inserting Eqs.~(\ref{dT0}), (\ref{Q12_1})--(\ref{Q12_N}) in
Eq.~(\ref{Tmatrix_De}) I find at $r\rightarrow \infty$
%\begin{equation}
%	\label{dT1}
%	T(r) =	\begin{bmatrix}
%		T_{+}(r) &  	{\sqrt \imath (-1)^{L}} T_{+-}(r) \\
%		(1+(-1)^{L}\imath)T_{-+}(r) &  	\imath T_{-}(r)
%	\end{bmatrix},
%\end{equation}
%and at $r\rightarrow \infty$
\begin{equation}
	\label{dTfinal}
	T =	\begin{bmatrix}
		\tilde{ T}_{+} & \imath^{L+1/2}\tilde{ T}_{+-} \\
	\imath^{L+1/2}\tilde{ T}_{-+} &  	\imath \tilde{ T}_{-}
	\end{bmatrix},
\end{equation}
where again  submatrices $	 \tilde{ T}_{+}$, $	 \tilde{ T}_{+-}$, $	 \tilde{ T}_{-}$ are real; and $ \tilde{ T}_{-+}$ is a transposition of $ \tilde{ T}_{+-}$. In Eqs.~(\ref{dT0}), (\ref{dTfinal}) it is assumed that the parity of $L$ is preserved, otherwise the factor at $T_{+-}$  is more complicated, remaining a complex constant.%  equal in modulus to $1$.

From Eq.~(\ref{dTfinal}) it can be seen that the number of real independent parameters (functions of $q_{1}$) defining the $T$ and $S$ matrices is equal to $(n^2+n)/2$  (independent of $q_{1}$).
This result can be improved.

Inserting 	Eq.~(\ref{dTfinal} )  in Eq.~(\ref{SfromT_inf}) I get
\begin{eqnarray}
	\label{S_Nchannels}
	S =	\begin{bmatrix}
S_{+} &  S_{+-} \\
S_{-+}	 &  S_{-}
	\end{bmatrix},
\end{eqnarray}
where
\begin{eqnarray}
	\label{T_for_S_Nchannels}
	S_{+}=	(1-\imath T_{+})^{-1}(1+\imath T_{+}),\\
		S_{+-}= (1-\imath T_{+})^{-1} \imath 	\sqrt{ \imath (-1)^{L}}G,\\
		S_{-+}= S_{+-}^{T} = \imath \sqrt{ \imath (-1)^{L}}(1+A)^{-1}\tilde{ T}_{-+} (1+\tilde{ S}),\\
	S_{-}=		(1+A)^{-1}(1-A),\\
		 \tilde{ S}=(1-\imath \tilde{ T}_{+})^{-1}(1+\imath \tilde{ T}_{+}), \\
			G=\tilde{ T}_{+-}(1+	(1+ \tilde{ T}_{-})^{-1}(1-	 \tilde{ T}_{-})),\\
T_{+}=\tilde{ T}_{+}-(-1)^{L}\tilde{ T}_{+-}(1+\tilde{ T}_{-})^{-1}\tilde{ T}_{-+},\\
A=\tilde{ T}_{-}+\imath (-1)^{L}\tilde{ T}_{-+}(1-\imath \tilde{ T}_{+})^{-1}\tilde{ T}_{+-}.
\end{eqnarray}
The $S_{-}$ submatix can be written as
\begin{eqnarray}
	\label{Sm_nchannels}
S_{-}=S_{-+}(1+S_{+})^{-1}S_{+-}+\tilde{\Delta},\\
\text{where } \tilde{\Delta}=	(1-\tilde{T}_{-})(1+\tilde{T}_{-})^{-1}.
\end{eqnarray}

Substituting Eqs.~(\ref{Q12_1}), (\ref{rho_my}), and (\ref{Q12_H}) into Eq.~(\ref{f4first}) and applying the relation $S(-q_{1})=S(q_{1})^{*}$, I find that $F(x,y)$ is independent of the real matrix $\tilde{\Delta}$. Therefore, when determining the potential from Eqs.~(\ref{Vpot6}), I may set
\begin{eqnarray}
	\label{Sm_nchannels_cut}
	S_{-}=S_{-+}(1+S_{+})^{-1}S_{+-}.
\end{eqnarray}
Consequently, the number of real parameters (functions of $q_{1}$) required to define a "sufficient" for IP solution $S$-matrix is
\begin{equation}
\frac{n_{+}^2+n_{+}}{2}+n_{+}(n-n_{+})=\frac{n_{+}(1-n_{+}+2n)}{2}. 
\end{equation}
Here, the $(n_{+}^2+n_{+})/2$ parameters specify the open-channel submatrix $S_{+}$ via a real, symmetric $T_{+}$ matrix of dimension $n_{+}\times n_{+}$, while the $n_{+}(n-n_{+})$ parameters specify the coupling submatrix $S_{+-}=S^{T}_{-+}$ via the same $T_{+}$ together with a real $G$ matrix of dimension $n_{+}\times(n-n_{+})$.

For the two-channel case below $W_{2}$, I obtain
\begin{eqnarray}
	\label{S_2channels_below}
	S =	\begin{bmatrix}
		e^{	2\imath\delta_{1}} &  \imath^{L_{2}+3/2}	e^{\imath\delta} \gamma\\
		\imath^{L_{2}+3/2}	e^{\imath\delta} \gamma &  	\frac{(-1)^{L_{2}+1}\imath \gamma^2}{	e^{	2\imath\delta}+1 }+\phi
	\end{bmatrix},\\
	S_{22}=S_{12}^2/(S_{11}+1)+\phi.
\end{eqnarray}
Here, $\delta_1$, $\gamma$, and $\phi$ are real functions of $q_{1}$, with $\phi$ being  arbitrary (e.g., $\phi=0$). Above $W_2$, the standard parametrization applies:
\begin{eqnarray}
	\label{S_2channels_above}
	S =	\begin{bmatrix}
		e^{	2\imath\delta_{1}}\cos(2\varepsilon) & \imath	e^{	\imath(\delta_{1}+\delta_{2})}\sin(2\varepsilon)    \\
		\imath	e^{	\imath(\delta_{1}+\delta_{2})}\sin(2\varepsilon) &  		e^{	2\imath\delta_{2}}\cos(2\varepsilon) 
	\end{bmatrix},
\end{eqnarray}
where  $\delta_1$, $\delta_2$,  and $\varepsilon$ are real functions of $q_{1}$.
It is assumed that functions $\delta_1$, $\delta_2$,  and $\varepsilon$ can be obtained from experimental data.  The parameter $\gamma$, however, cannot be determined from scattering data; the possibility of  its experimental determination requires further research.

To analyze the $T$-matrix behavior at the threshold $W_{j}$ ($q_j=0$)  it is convenient to consider Eq.~(\ref{Tmatrix_De}). Proceeding analogously to the derivation of Eq.~(\ref{dTfinal}) but employing the small-argument expansions $J(Qr)_{jj}\propto (q_{j}r)^{L+1}$ and  $N(Qr)_{jj}\propto (q_{j}r)^{-L}$ I obtain the relations 
\begin{equation}
T_{kj} \xrightarrow[q_{j} \to \imath 0; 0]{}	\begin{cases}
	b_{kj}q_{j}^{L_{j}+1/2}, & k \neq j \\
	a_{kj}q_{j}^{2L_{j}+1}	 & k=j
\end{cases}
\label{T_at_thres}	, 
\end{equation} 
 with real constants $a_{kj}$ and $b_{kj}$. The constants $a_{kj}$ take the same value whether the threshold is approached from the imaginary momentum axis ($q_j = i|q_j| \to i0$) or from the positive real axis ($q_j = |q_j| \to +0$).
 
For the two-channel case, the behavior of $\gamma$ below $W_2$ for small imaginary $q_2 = i|q_2|$ can be determined from Eqs.~(\ref{SfromT_inf}) and (\ref{T_at_thres}):
\begin{equation}
\gamma \xrightarrow[q_{2} \to \imath 0]{} 2b \cos(\delta_1)|q_{2}|^{L_{2}+1/2}
	\label{gamma_at_thres_left}	, 
\end{equation} 
 where $\delta_1$ is continuous across the threshold, and the constant $b$ is fixed by the limit above $W_2$ (for $q_2 = |q_2| \to +0$):
 \begin{equation}
 	\sin (2\varepsilon) \xrightarrow[q_{2} \to +0]{} 2b \cos(\delta_1)q_{2}^{L_{2}+1/2}
 	\label{gamma_at_thres_right}	. 
 \end{equation}
A similar analysis yields 
  \begin{equation}
  	\gamma \xrightarrow[q_{1} \to +0]{} c q_{1}^{L_{1}+1/2}
  	\label{gamma_at_q1=0}	.
  \end{equation} 
 %In this paper I consider a case of only two coupled channels, and $m_{11}=m_{21}=m_{N}$, $m_{12}=m_{\pi}$, $m_{22}=m_{\eta}$  are nucleon, $\pi$, and $\eta$ masses correspondingly.

\section{An approximation of the $S$ matrix}
This paper tests and applies the developed formalism to the two-channel case; the detailed implementation is described below.
The method of \cite{MyAlg4} admits a straightforward generalization to the coupled-channel case. As a first approximation, I employ the threshold-free $S$-matrix given by Eq.~(\ref{S_2channels_above})
\begin{widetext}
 \begin{eqnarray}
	\label{eqS2}
	\begin{array}{l}
		S^{(0)}(q) 
		= \left( {{\begin{array}{*{20}c}
					{\left( {\frac{f_2^{\left( 1 \right)} \left( q \right) + \imath f_1^{\left( 1
									\right)} \left( q \right)}{f_2^{\left( 1 \right)} \left( q \right) - \imath  f_1^{\left( 1 \right)} \left( q \right)}}
						\right)^2\frac{\left( {f_2^{\left( {12} \right)} \left( q \right)} \right)^2 - \left( {f_1^{\left( {12} \right)} \left(
								q \right)} \right)^2}{\left( {f_2^{\left( {12} \right)} \left( q \right)} \right)^2 + \left( {f_1^{\left( {12} \right)}
								\left( q \right)} \right)^2}} \hfill
					& {  \frac{- 2 \imath f_2^{\left( {12} \right)} \left( x \right)f_1^{\left( {12} \right)} \left( x \right)}{\left(
							{f_2^{\left( {12} \right)} \left( q \right)} \right)^2 + \left( {f_1^{\left( {12} \right)} \left( q \right)}
							\right)^2}\prod\limits_{j = 1,2} {\frac{f_2^{\left( j \right)} \left( q \right) + \imath f_1^{\left( j \right)} \left( q
								\right)}{f_2^{\left( j \right)} \left( q \right) - \imath f_1^{\left( j
									\right)} \left( q \right)}} } \hfill \\
					{ \frac{- 2 \imath f_2^{\left( {12} \right)} \left( q \right)f_1^{\left( {12}
								\right)} \left( q \right)}{\left( {f_2^{\left( {12} \right)} \left( q \right)} \right)^2 + \left( {f_1^{\left( {12}
									\right)} \left( q \right)} \right)^2}\prod\limits_{j = 1,2} {\frac{f_2^{\left( j \right)} \left( q \right) + \imath
								f_1^{\left( j \right)} \left( q \right)}{f_2^{\left( j \right)} \left( q \right) - \imath f_1^{\left( j \right)} \left( q
								\right)}} }  & {\left( {\frac{f_2^{\left( 2 \right)} \left( q \right) + \imath f_1^{\left( 2 \right)} \left( q
								\right)}{f_2^{\left( 2 \right)} \left( q \right) - \imath f_1^{\left( 2 \right)} \left( q \right)}} \right)^2\frac{\left(
							{f_2^{\left( {12} \right)} \left( q \right)} \right)^2 - \left( {f_1^{\left( {12} \right)} \left( q \right)}
							\right)^2}{\left( {f_2^{\left( {12} \right)} \left( q \right)} \right)^2 + \left( {f_1^{\left( {12} \right)} \left( q
								\right)} \right)^2}} \hfill \\
		\end{array} }} \right) \\
	\end{array}
 \end{eqnarray}
\end{widetext}
Here, $f_1^{(\cdot)}$ and $f_2^{(\cdot)}$ are odd and even polynomials in $q$, respectively, and they do not vanish simultaneously.
From here on $q\equiv q_{1}$.
This rational approximation has been employed previously to describe $NN$ scattering \cite{Funk2001, Geramb1994, Kohlhoff1994, Khokhlov2006, Khokhlov2007}.    The polynomial coefficients are real and are determined by the following relations:
\begin{eqnarray}
	\label{eq6b} \tan \left( \frac{\delta_i(q)}{2} \right) = - \frac{f^{(i)}_1 (q)}{f^{(i)}_2 (q)}, \quad i=1,2,\ \\
	\label{eq13} \tan \left( \varepsilon(q) \right) = -\frac{f^{(12)}_1 (q)}{f^{(12)}_2 (q)}.
\end{eqnarray}
To obtain these coefficients, one specifies the $S$-matrix elements at a set of momentum values corresponding to the number of unknowns. The coefficients are then found by solving the resulting system of linear equations.
Despite the apparent simplicity of this algorithm, it shares similar pitfalls with the single-channel case. 
Namely, increasing the accuracy requires more coefficients, which in turn can cause artificial poles of the $S$ matrix to appear uncontrollably near or on the real $q$ axis.
In our case, the situation is further complicated. The matrix given by Eq.~(\ref{eqS2}) is unitary for all real $q$, but the $S$ matrix in the problem under consideration must be non-unitary below the threshold $W_{2}$. Specifically, below $W_{2}$ we have $S_{11} = \exp(2 i \delta_{1})$ with real $\delta_{1}$. However, as in the single-channel case, this expression can serve as a rough approximation of the $S$-matrix above $W_{2}$. In particular, the degrees of the polynomials in Eq.~(\ref{eqS2}) can be chosen to correctly describe the $S$-matrix asymptotics at large $q$, $\delta_i, \varepsilon \propto 1/q$.
As noted previously, selecting appropriate coefficients for Eq.~(\ref{eqS2}) presents a significant challenge; despite methodological advances \cite{no_poles_methods}, a fully reliable algorithm remains elusive. Here, we adopt an approach analogous to the earlier single-channel case. The polynomial degrees are gradually increased, while the set of momentum points $q$ above threshold—at which the $S$ matrix is fixed to experimental values (i.g., via a spline fit)—is varied in a quasi-random manner. Once a certain limiting degree is reached (typically around 10 for  experimental data), avoiding the appearance of artificial poles becomes virtually impossible. This best approximation—free of artificial poles—is subsequently refined in the next step.

The initial approximation is subsequently refined by adding a truncated sinc series to each element of the $S$-matrix. Using the formulas below, the series coefficients are determined from the target $S$ matrix $S_{e}(q)$, ensuring its reproduction by the IP solution.

Let $R$ be the interaction range, and let $2R\tau =\pi$. The set of sinc functions with approximately finite support and an approximately separable Fourier transform is defined by
\begin{eqnarray}
	\lambda_{k}(q)=\frac{\sin\left((k\tau-q)\pi/\tau\right)}{(k\tau-q)\pi/\tau}. \label{sin_basis_L0}
\end{eqnarray}

Next, I introduce the deviation
\begin{equation}
	\Delta S^{(j)} = S_{e}(q^{(j)}) - S^{(0)}(q^{(j)}),
	\label{dSmatrixN3}
\end{equation}
where $q^{(j)},\ j=1,\dots,N$ are the momentum values for which a description of the target $S$ matrix is required.
I interpolate these deviation values using a second-order spline, denoted $\Delta_{2} S(q)$, such that $\Delta_{2} S(q^{(j)})= \Delta S^{(j)}$. This spline is then used to calculate
\begin{equation}
	\Delta S(q)=- \sum_{k=-N}^{N} s_{k} \lambda_{k}(q)\approx \Delta_{2} S(q).
	\label{SmatrixN3}
\end{equation}
The matrix coefficients $s_{k}$ are determined by
\begin{equation}
	s_{k} = -\left.\Delta_{2} S(q)\right|_{q=k\tau},\
	s_{0}\equiv 0.
	\label{s_k}
\end{equation}
    The result is a sufficiently accurate approximation of the $S$ matrix, whose precision is governed by the parameters $R$ and $N$. This approximation is free of unphysical poles and yields an approximately separable input kernel for the Marchenko equation:
\begin{equation}
	S(q)=S^{(0)}(q)- \sum_{k=-N}^{N} s_{k} \lambda_{k}(q)\approx S_{e}(q).
	\label{final_Smatrix_ap}
\end{equation}
\begin{widetext}
Using this approximation, I follow the established procedure \cite{Funk2001,Geramb1994,Kohlhoff1994,Khokhlov2006,Khokhlov2007,MyAlg4} by substituting Eq.~(\ref{final_Smatrix_ap}) into Eq.~(\ref{f4first}). With approximations of \cite{MyAlg4}, this yields 
\begin{multline}
	F\left( {x,y} \right)\\
	 \approx \imath \sum\limits_{j=1}^{n_{pos}}
\textrm{Res}
 \left[	H^{(+)}(Qx)\rho \left( {I - S^{(0)}\left( q
	\right)} \right)\rho	H^{(+)}(Qy) \right]_{q=\beta_{j}} 
		+\sum\limits_{j = 1}^{n_{\mbox{\tiny{b}}} } {	H^{(+)}(\tilde{Q}_{j} x)  A_j 	H^{(+)}(\tilde{Q}_{j} y) }\\
				+\frac{\tau}{2 \pi}\sum\limits_{k = -N}^{N} 
							H(R-|x|) \left[ H^{(+)}(Qx) 
				q\rho	s_{k} \rho
				H^{(+)}(Qy)\right]_{q=k\tau} 	H(R-|y|)  \\
		%%%%%%%%%%%%%%%%%%%%%%%%%%%%%%%%%%%%%%%%%%%%%%%%%%%%%%%%%%
	 \approx \sum\limits_{j=1}^{M} 	\Lambda_{j}(x)\Omega_{j}	\Lambda_{j}(y)+
	 \sum\limits_{j=M+1}^{M+n_{pos}^{(2)}} 	{\tilde \Lambda}_{j}(x)\Omega_{j}	\Lambda_{j}(y)
	 +\sum\limits_{j=M+n_{\mbox{\tiny{pos}}}^{(2)}+1}^{M+2n_{\mbox{\tiny{pos}}}^{(2)}} 	\Lambda_{j}(x)\Omega_{j}	{\tilde \Lambda}_{j}(y),
	 \label{input_sep}
\end{multline}
\end{widetext}
where $H(z)$ is the Heaviside step function and $M=n_{pos}+2N+n_{b}$.
The functions $\Lambda_j(x)$ are defined in two groups. The first group is given by
\begin{equation}
	\Lambda_j \left( { x} \right) =
		H(R-|x|) H^{(+)}(Qx)\big|_{q=\beta_{j}},
	\end{equation}
	for $j = 1,\dots,M$, where $M = n_{\text{pos}} + 2N + n_b$. Here, the parameters $\beta_j$ are grouped as follows:
		\begin{itemize}
		\item 	$j = 1,\dots,n_{\text{pos}}$: all $S$-matrix poles (first and second order) with $\text{Im}[\beta_j] > 0$;
		\item $j = 1+n_{\text{pos}},\dots,n_{\text{pos}}+n_b$: imaginary momenta corresponding to bound states;
		\item $j = 1+n_{\text{pos}}+n_b,\dots,n_{\text{pos}}+n_b+2N$: $\beta_j = k\tau$, where $k = -N,\dots,N$.
	\end{itemize}
The second group involves the derivative of $H^{(+)}$ and is defined as follows:
\begin{eqnarray}
\tilde{	\Lambda}_j(x) = x H(R-|x|) \dot H^{(+)}(Qx)\big|_{q=\beta_j},\\
{	\Lambda}_j(x) = H(R-|x|)  H^{(+)}(Qx)\big|_{q=\beta_j}
\end{eqnarray}
for $j = M+1,\dots,M+2n_{\text{pos}}^{(2)}$, where
	\[
{\dot H}^{(+)}(x) = \left( {{\begin{array}{*{20}c}
			{{dh_{l_1 }^ + \left( x \right)} \mathord{\left/ {\vphantom {{dh_{l_1 }^ +
								\left( x \right)} {dx}}} \right. \kern-\nulldelimiterspace} {dx}}
			\hfill & 0
			\hfill \\
			0 \hfill & {{dh_{l_2 }^ + \left( x \right)} \mathord{\left/ {\vphantom
						{{dh_{l_2 }^ + \left( x \right)} {dx}}} \right.
					\kern-\nulldelimiterspace}
				{dx}} \hfill \\
\end{array} }} \right).
\]
The parameters $\beta_j$ in this index range correspond to the second-order $S$-matrix poles. Specifically, the first $n_{\text{pos}}^{(2)}$ values in this range ($j = M+1,\dots,M+n_{\text{pos}}^{(2)}$) and the next $n_{\text{pos}}^{(2)}$ values ($j = M+n_{\text{pos}}^{(2)}+1,\dots,M+2n_{\text{pos}}^{(2)}$) are both the set of second-order poles.

The input kernel of the Marchenko equation is approximated by the manifestly separable form in Eq.~(\ref{input_sep}). The equation is then solved following \cite{Khokhlov2006} by means of the substitution
\begin{equation}
L(x,y)=\sum_{j=1}^{M}N_{j}(x)	\Lambda_j \left( { y} \right)
+\sum_{j=M+1}^{M+n_{\mbox{\tiny{pos}}}^{(2)}}P_{j}(x)	
{\tilde\Lambda}_j \left( { y} \right),
\end{equation}
with linear independent $\Lambda_j \left( { y} \right)$ and ${\tilde\Lambda}_j \left( { y} \right)$. 
The functions $\Lambda_j(y)$ appearing in the second sum of Eq.~(\ref{input_sep}) are already present in the first sum. Hence, the two sums can be combined.
This substitution leads to a system of linear equations for the matrix coefficients $N_{j}(x)$ and $P_{j}(x)$, which is solved for each $x$ where the potential is to be determined. The derivatives of the functions $\Lambda_j(y)$ and $\tilde\Lambda_j(y)$ are then found analytically. To obtain the derivatives of $N_{j}(x)$ and $P_{j}(x)$, Eq.~(\ref{f3}) is differentiated, yielding an integral equation with a separable kernel. This equation is solved using the same substitution. The potential is then calculated from Eq.~(\ref{Vpot6}).

\section{Results and Conclusions}
\begin{figure}[h]
	% Use the \centerline and \includegraphics commands to insert your figure file:
	\centerline{\includegraphics[width=0.42\textwidth]{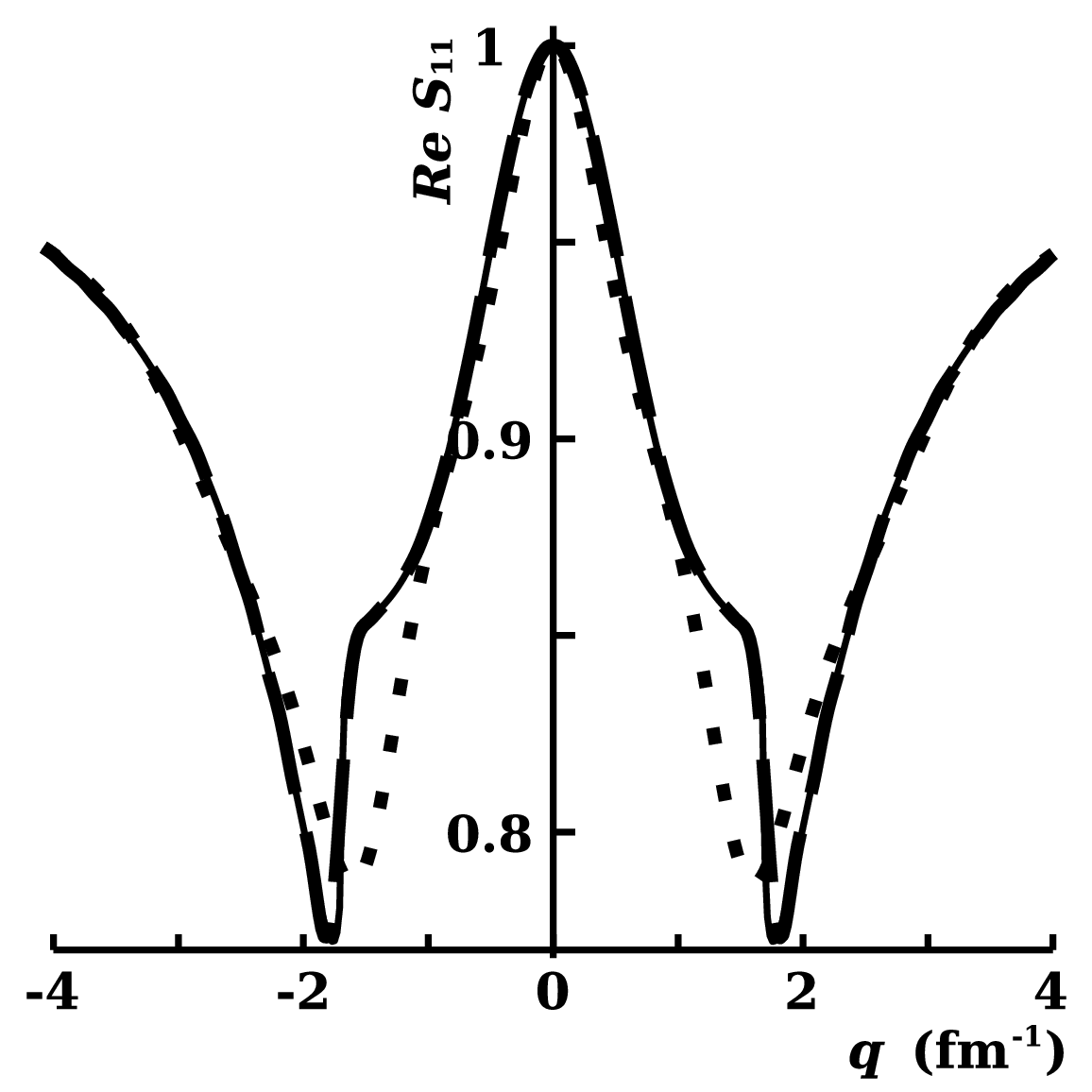}}
		\centerline{\includegraphics[width=0.42\textwidth]{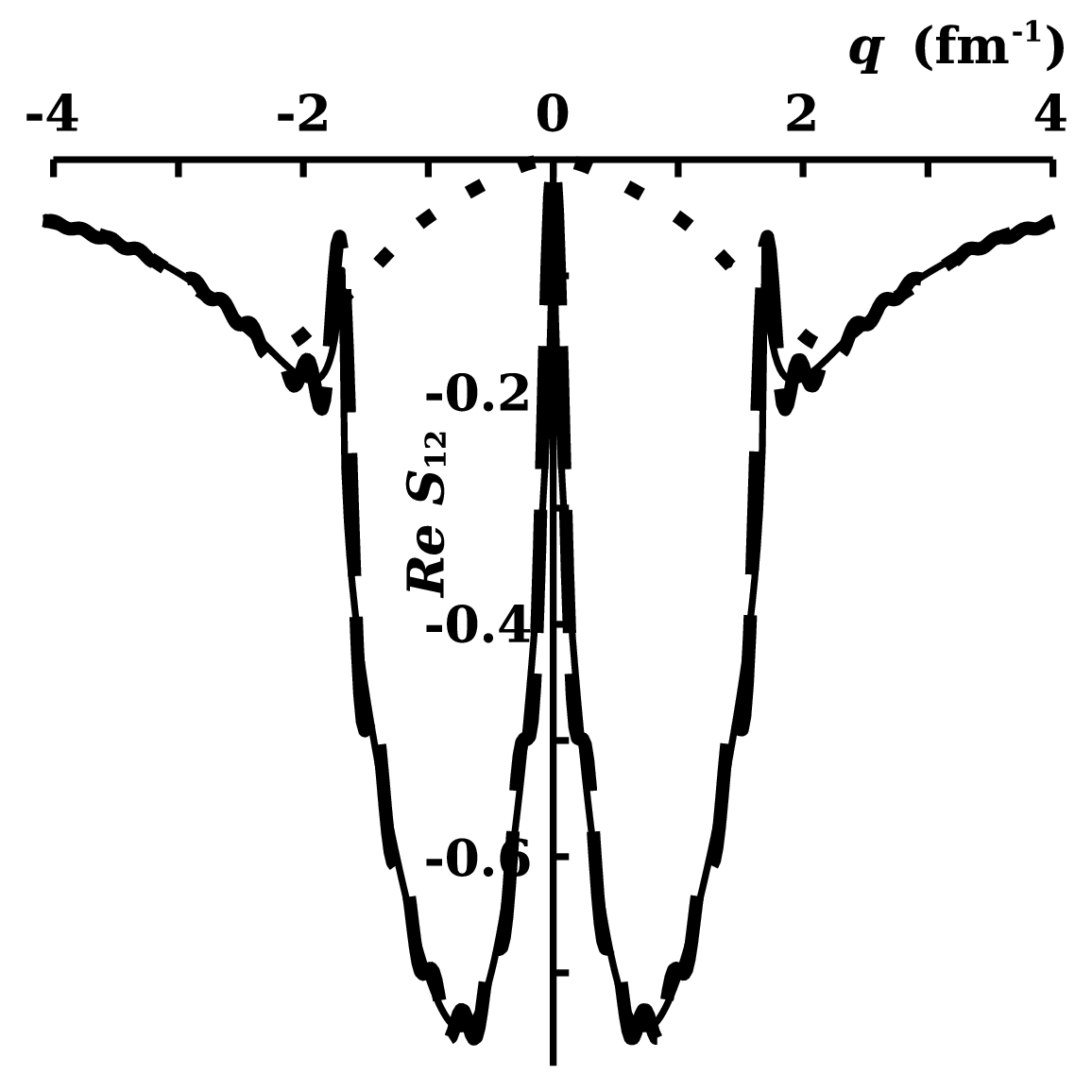}}
			\centerline{\includegraphics[width=0.42\textwidth]{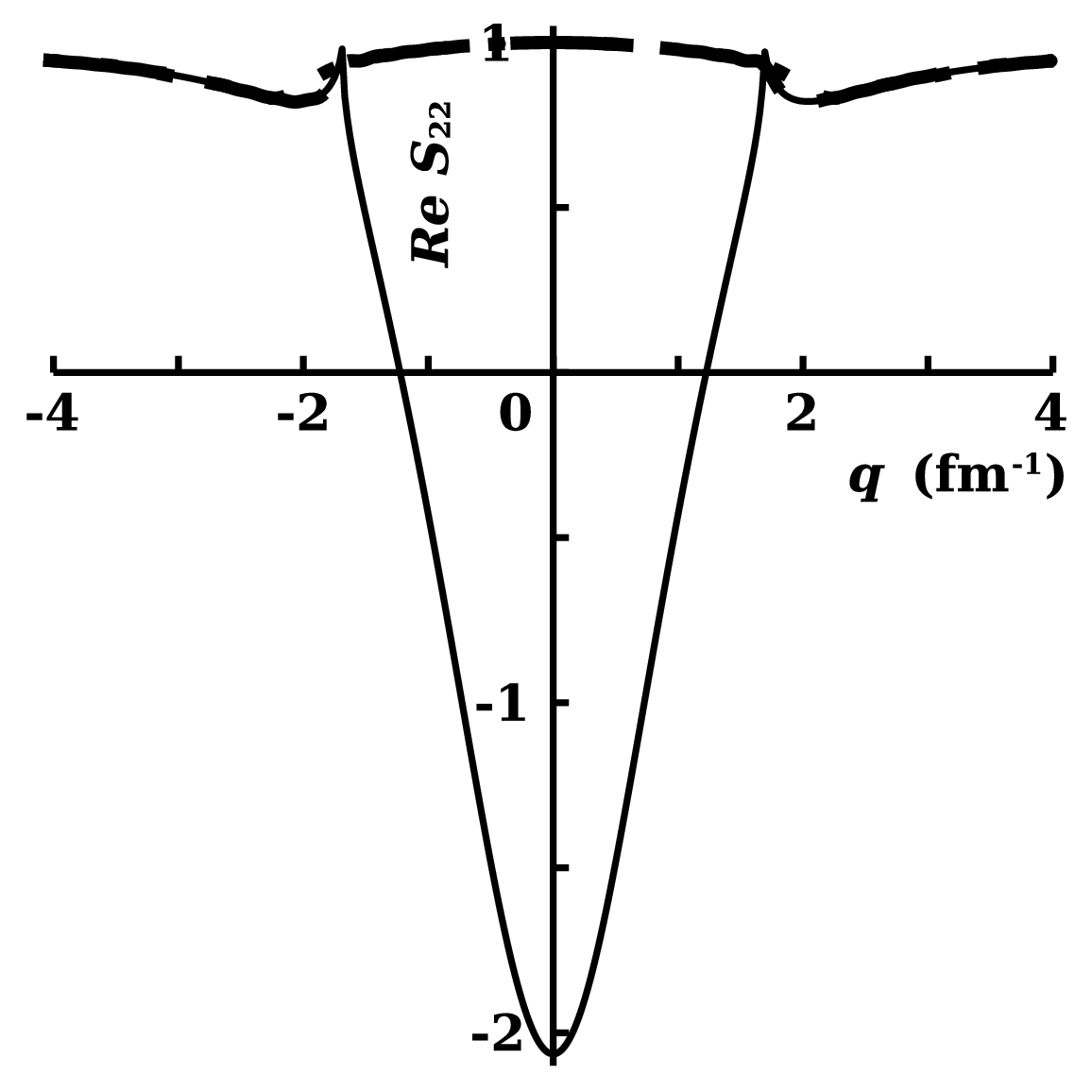}}
	\caption{\label{fig:testdataRe} The real parts of the $S$-matrix elements used to reconstruct the test potential $V_{1}(r)=V_{2}(r)=V_{12}(r)=V_{0}e^{-ar^{2}}$, with $V_{0}=-1.5\ \text{fm}^{-2}$ and $a=2\ \text{fm}^{-2}$, are shown. The solid thin curves were obtained from the direct solution of Eq.~(\ref{system2by2}), while the thick dashed curves represent the approximation $S(q)=S^{(0)}(q)+\Delta S(q)$, and the dotted curves correspond to $S^{(0)}(q)$ alone.
			}
\end{figure}

\begin{figure}[h]
	% Use the \centerline and \includegraphics commands to insert your figure file:
	\centerline{\includegraphics[width=0.45\textwidth]{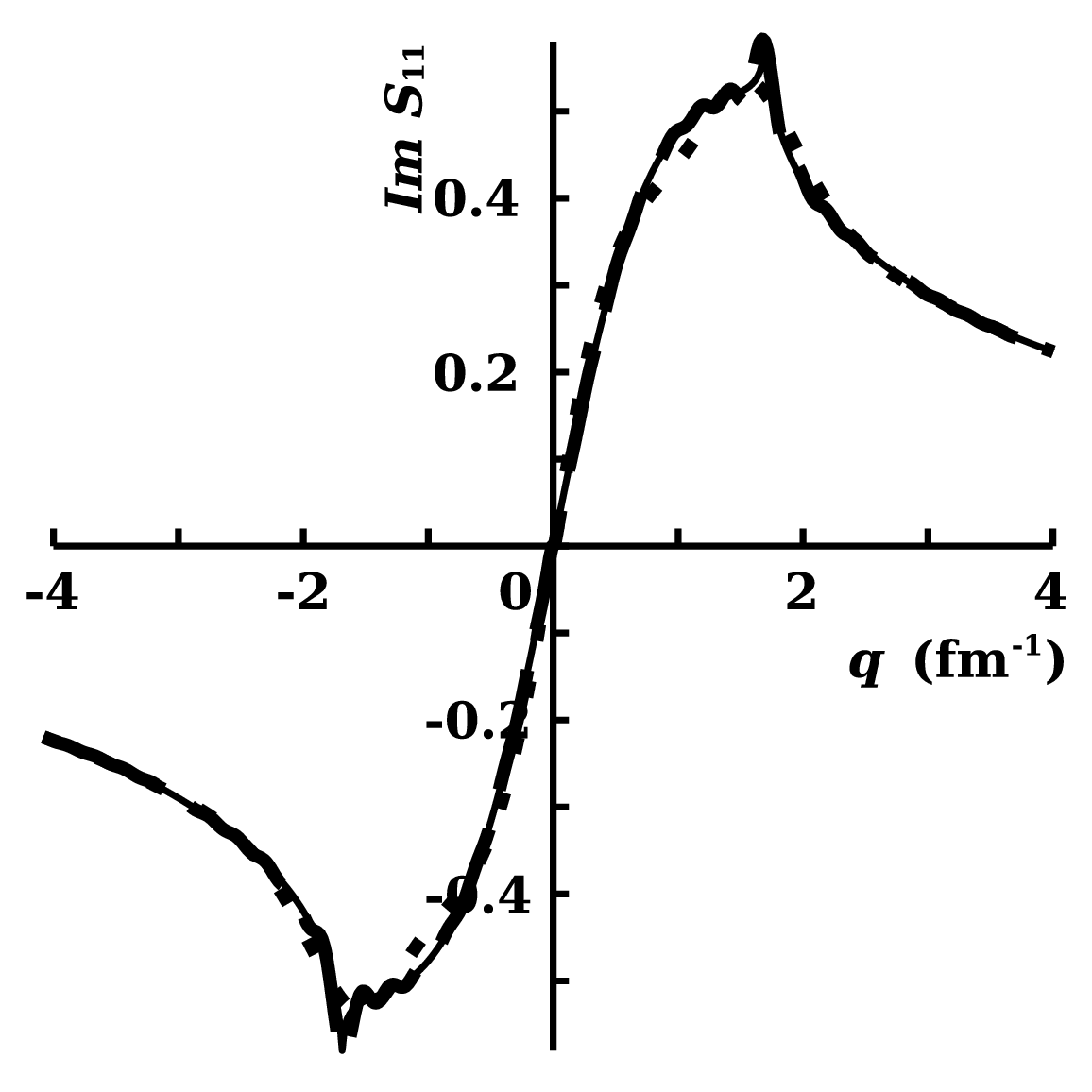}}
	\centerline{\includegraphics[width=0.45\textwidth]{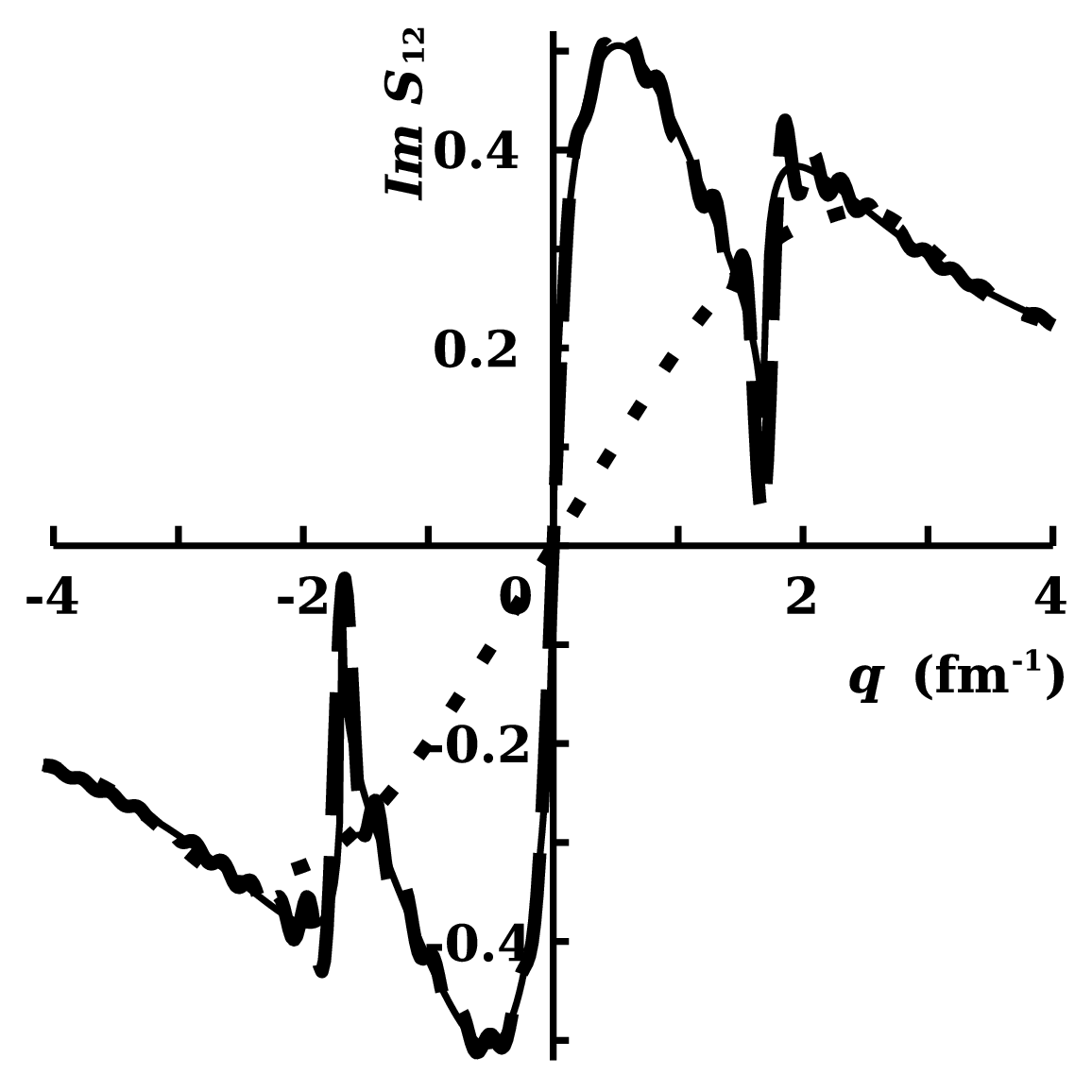}}
		\centerline{\includegraphics[width=0.45\textwidth]{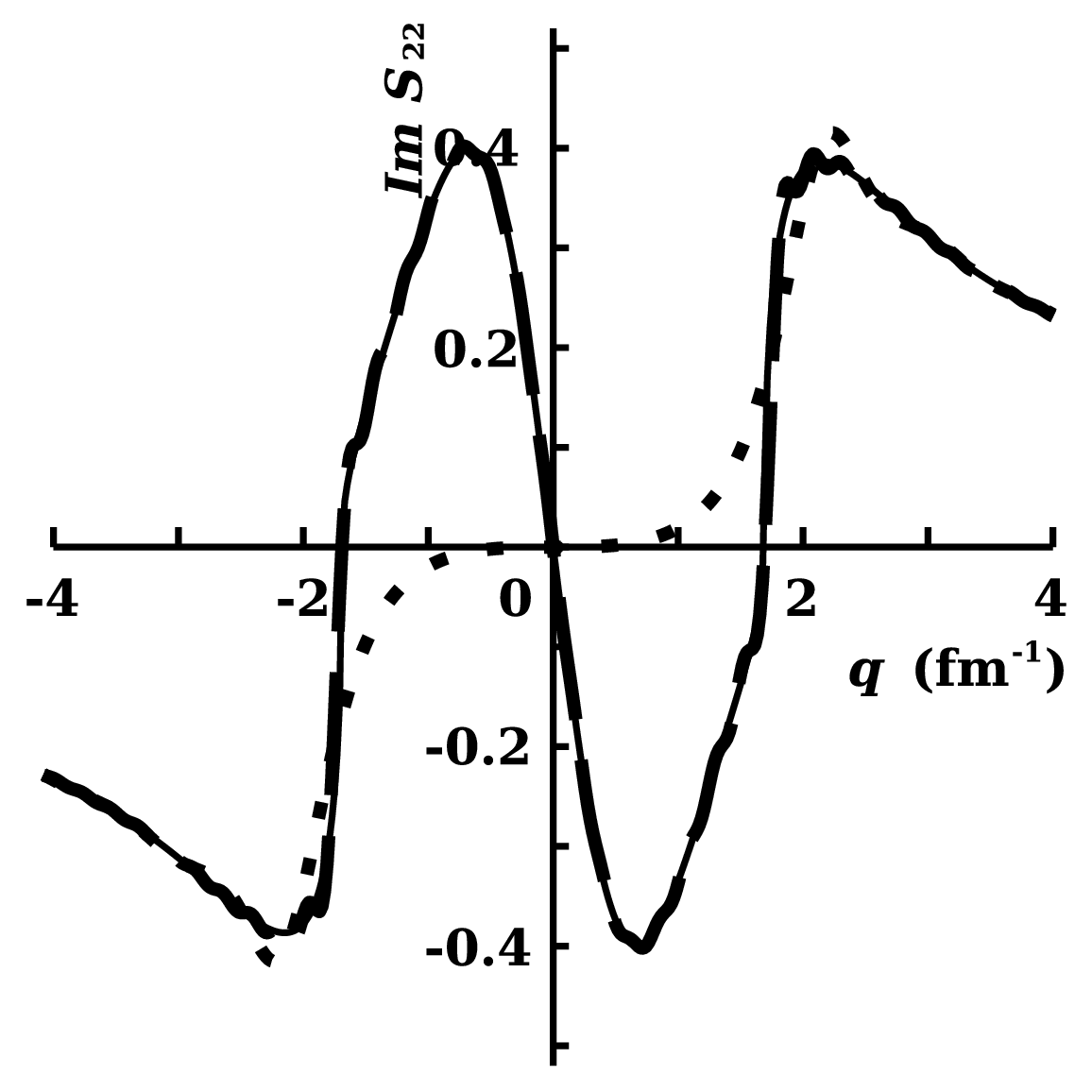}}
\caption{\label{fig:testdataIm} Continuation of Fig.~\ref{fig:testdataRe} for imaginary parts of the $S$-matrix elements.
	}
\end{figure}

\begin{figure}[h]
	% Use the \centerline and \includegraphics commands to insert your figure file:
	\centerline{\includegraphics[width=0.5\textwidth]{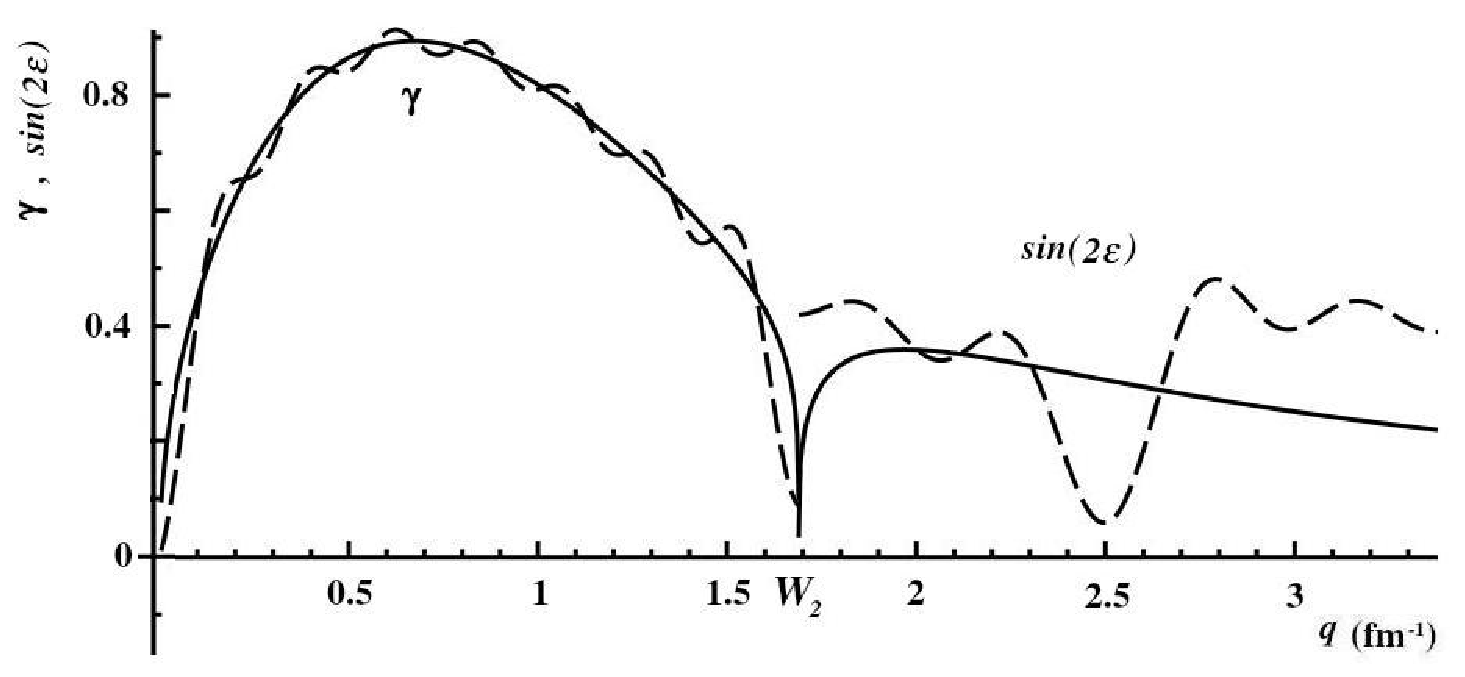}}
	%\centerline{\includegraphics[width=0.8\textwidth]{dwaves.eps}}
	% Use the \caption command to produce the figure caption and place it below the graph:
	\caption{\label{fig:exp_gamma} $S$ matrix parameters $\gamma$ (bellow $W_2$) and $\sin(2\epsilon)$ (above $W_2$).
	Solid  curves were calculated from direct solution of Eq.~(\ref{system2by2}), while dashed curves represent approximation Eq.~(\ref{final_Smatrix_ap}). }
\end{figure}

\begin{figure}[h]
	% Use the \centerline and \includegraphics commands to insert your figure file:
	\centerline{\includegraphics[width=0.5\textwidth]{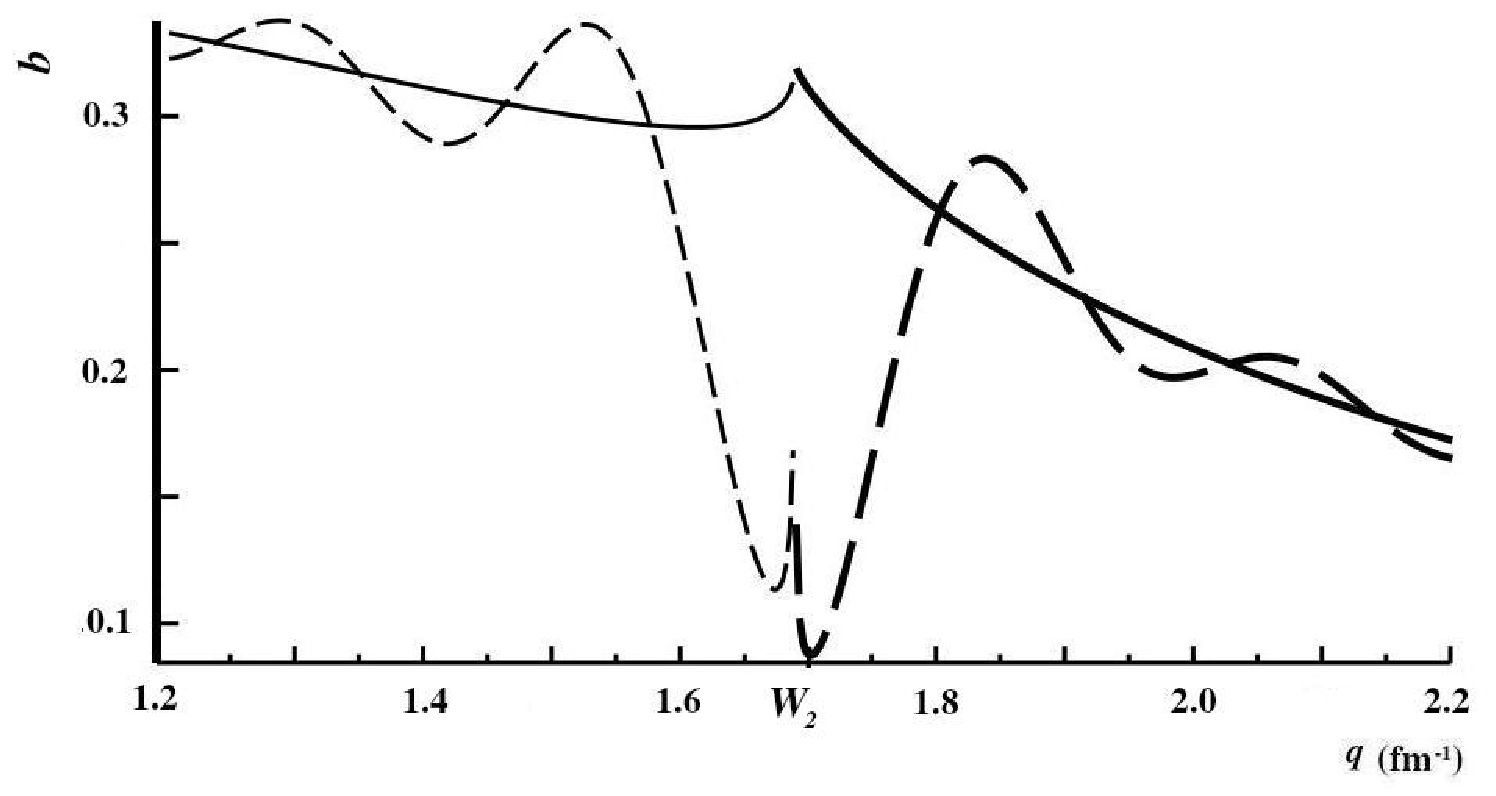}}
	%\centerline{\includegraphics[width=0.8\textwidth]{dwaves.eps}}
	% Use the \caption command to produce the figure caption and place it below the graph:
	\caption{\label{fig:exp_gamma_koef} Parameter $b$ defined from 
		Eq.~(\ref{gamma_at_thres_left}) bellow $W_2$ and from Eq.~(\ref{gamma_at_thres_right}) above $W_2$.
		Solid  curves were calculated from direct solution of Eq.~\ref{system2by2}, while dashed curves represents approximation Eq.~(\ref{final_Smatrix_ap}). }
\end{figure}

\begin{figure}[h] 
	% Use the \centerline and \includegraphics commands to insert your figure file:
	\centerline{\includegraphics[width=0.45\textwidth]{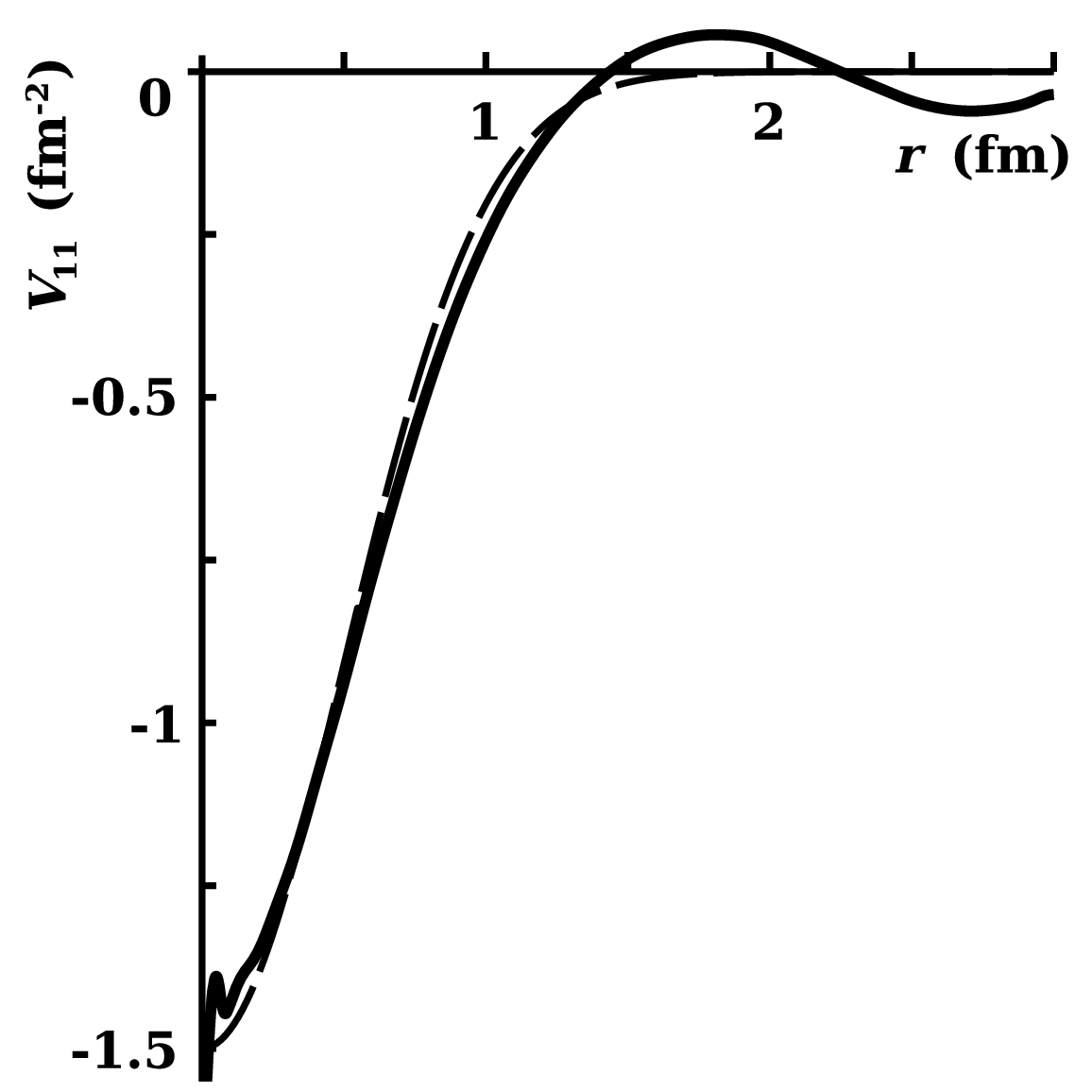}}
	\centerline{\includegraphics[width=0.45\textwidth]{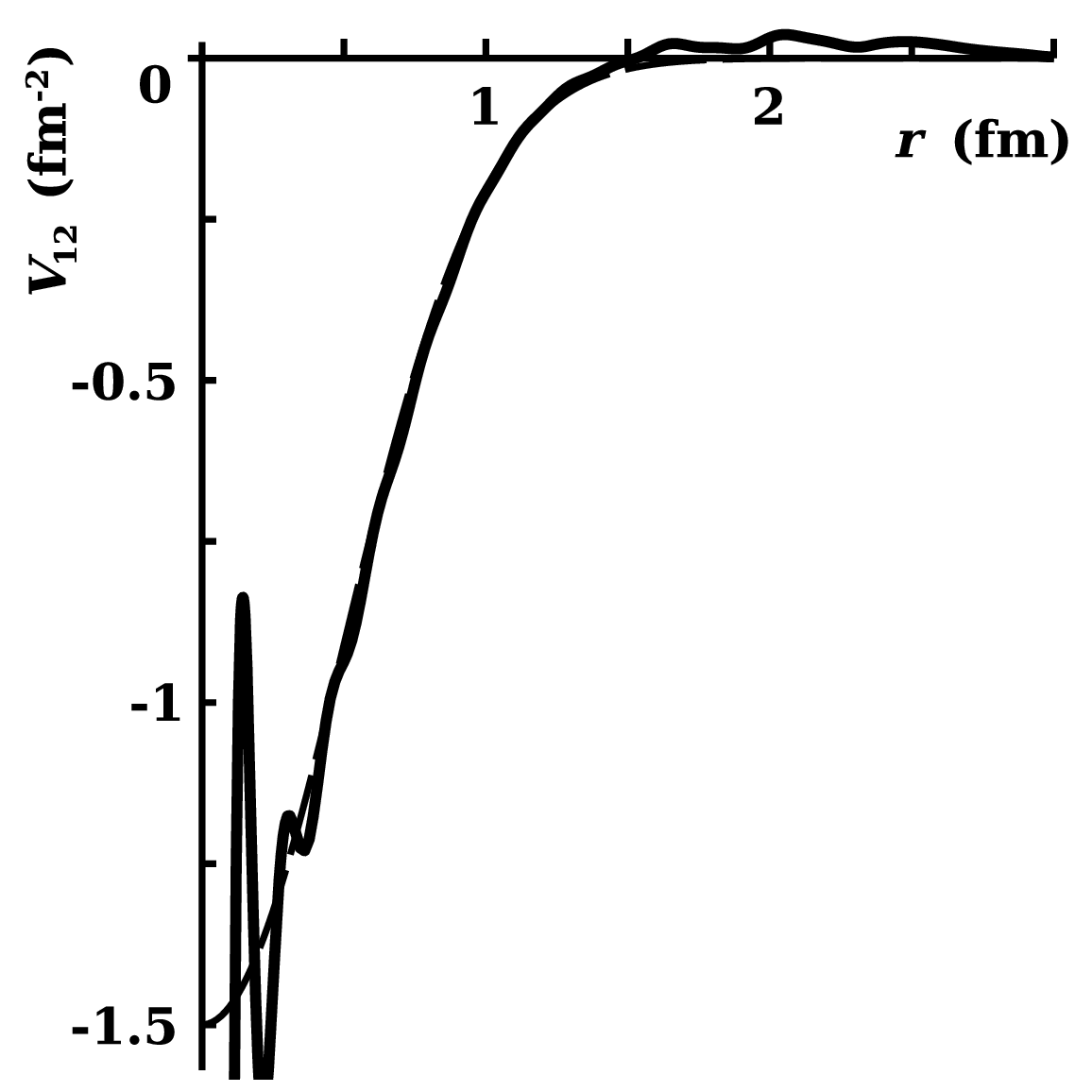}}
	\centerline{\includegraphics[width=0.45\textwidth]{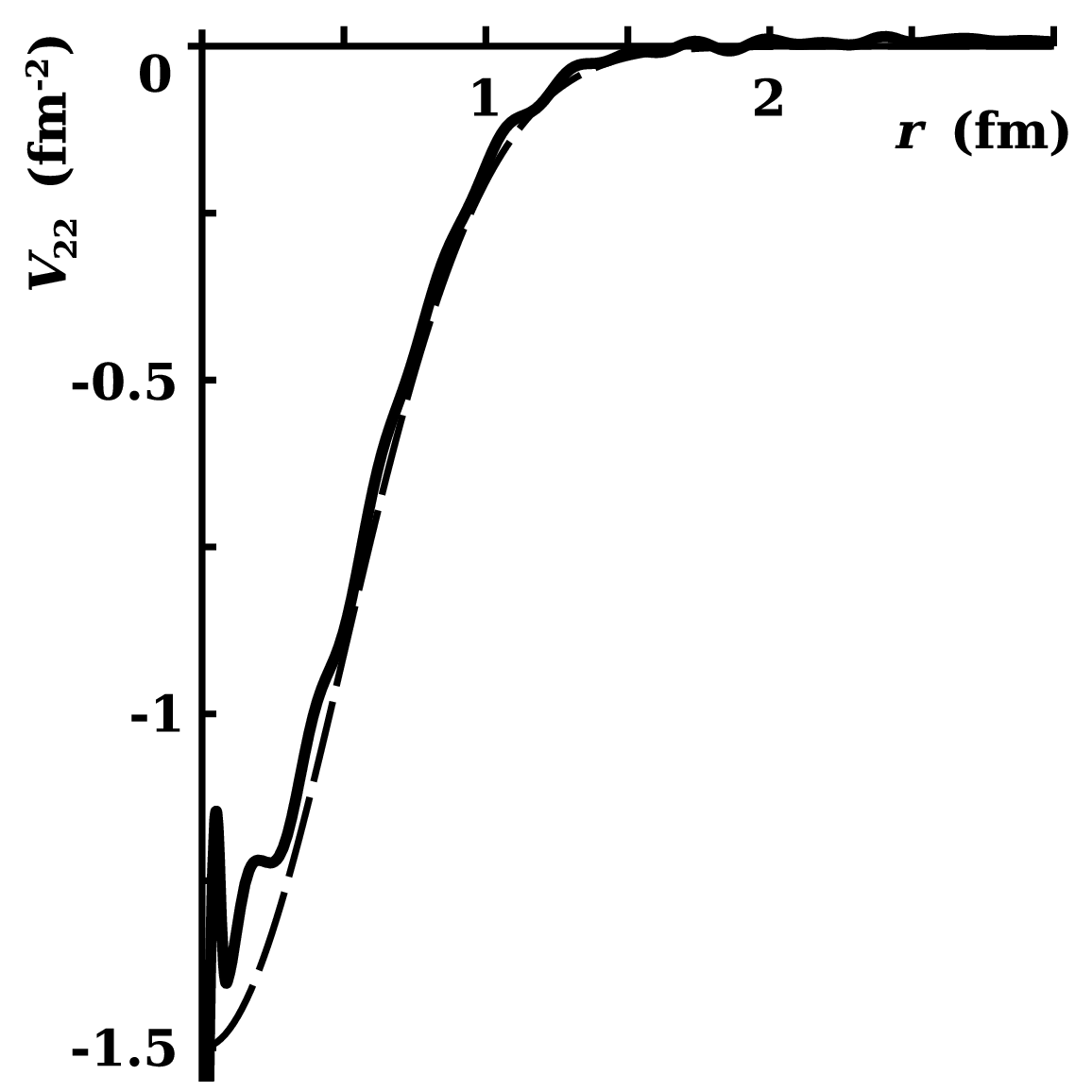}}
	%\centerline{\includegraphics[width=0.8\textwidth]{dwaves.eps}}
	% Use the \caption command to produce the figure caption and place it below the graph:
	\caption{\label{fig:example_pot_pit}
	Results of the inversion for the $S$-matrix fit presented in 
	Figs.~\ref{fig:testdataRe}-\ref{fig:exp_gamma_koef}. The dashed line corresponds to the input potential, and solid curves are results of the inversion. }
\end{figure}

The formalism and algorithm were tested on a number of two-channel threshold problems. Figs.~\ref{fig:testdataRe},~\ref{fig:testdataIm},~\ref{fig:exp_gamma},~\ref{fig:exp_gamma_koef},~\ref{fig:example_pot_pit} show the results for one representative example.
    In this problem, the elements of the $S$ matrix shown in Figs.~\ref{fig:testdataRe},~\ref{fig:testdataIm} were first obtained by numerically solving the direct problem for the two-channel case of Eq.~(\ref{system2by2}) with 
    \begin{equation}
    	L=0,\ V_{1}(r)=V_{2}(r)=V_{12}(r)=V_{0}e^{-ar^2},
    \end{equation}
    	  where $V_{0}=-1.5\ fm^{-2}$, $a=2\ fm^{-2}$.
   The particle masses in the channels were chosen as follows:
    	 \begin{multline}
    	 	m_{11}=m_{N}=939\text{ MeV}\approx 4.75\text{ fm}^{-1},\\ m_{12}=m_{\pi}=139.6\text{ MeV}\approx 0.708\text{ fm}^{-1},\\
    	 	m_{21}=m_{N}=939\text{ MeV}\approx 4.75\text{ fm}^{-1},\\ m_{22}=3m_{\pi}=418.8\text{ MeV}\approx 2.12\text{ fm}^{-1} \nonumber
    	 \end{multline}
   Here $m_{N}$ is the nucleon mass and $m_{\pi}$ the pion mass.
 This choice corresponds to scattering in the $S_{31}$ $\pi N$ channel, where inelastic channels open at $M = m_{N} + 3m_{\pi}$, that corresponds to $q\approx 1.69$~fm.
  The input $S_{e}(q)$ matrix was calculated up to $q\approx 19.6$~fm$^{-1}$ with step $\delta q\approx 0.0491$~fm$^{-1}$. 
 This data (interpolated by a second-degree spline) was then used to find the initial rational approximation $S^{(0)}(q)$, with polynomial degrees 36, 44, and 36 for $S^{(0)}_{11}(q)$, $S^{(0)}_{12}(q)$, and $S^{(0)}_{22}(q)$, respectively. Next, the matrix coefficients $s_{k}$ of the correction term from Eq.~(\ref{s_k}) were calculated for $q\lesssim 13.95$~fm$^{-1}$. The basis set parameters in Eq.~(\ref{sin_basis_L0}) were chosen as $R\approx 13.95$~fm and $\tau=\pi/2R\approx 0.113$~fm$^{-1}$. 
 
 The accuracy of the approximation is illustrated in Figs.~\ref{fig:testdataRe},~\ref{fig:testdataIm},~\ref{fig:exp_gamma}, and~\ref{fig:exp_gamma_koef}. In particular, Figs.~\ref{fig:testdataRe} and~\ref{fig:testdataIm} show a significant improvement of the initial rational approximation by the correction term in and below the threshold region. 
%%%%%%%%%%%%%%%%%%%%%%%%%%%%%%%%%%%%%%%%%
In accordance with the conclusions of Sect.~IV, the values of $Re\,S_{22}$ below the threshold were chosen arbitrarily (taken as $Re^{(0)}\,S_{22}$). Figure~\ref{fig:exp_gamma} illustrates the conclusions of Sect.~IV regarding the behavior of the real $\gamma$ function near and below the threshold. Figure~\ref{fig:exp_gamma_koef} confirms the possibility of extracting information about the subthreshold behavior of the $S_{12}$ element from its behavior above the threshold.

Following the algorithm described above, the inverse problem was then solved. The results, presented in Fig.~\ref{fig:example_pot_pit}, show satisfactory convergence of the method. Some divergence near $r=0$~fm is observed, which is inherent in calculations performed within the framework of Marchenko theory. 
Minor oscillations are also present at large $r$, decreasing slowly toward $R$. These oscillations are an artifact of the rational approximation and occur when the approximation has poles too close to the real $q$ axis.

To test the developed formalism against experimental data, I conducted an analysis of PWA data for the $\pi N$ $S_{31}$ state \cite{TpiN_data}. For this state, data are available up to  $E_{\text{lab}} \approx 2.5$~GeV.

The $T$-matrix behavior shows no significant features up to about $E_{\text{lab}} \approx 0.8$~GeV. In this region, resonant behavior is observed, which after $E_{\text{lab}} \approx 1$~GeV changes back  to a relatively smooth energy dependence. Inelastic channels open at the threshold $W_{2} = m_{N} + 3m_{\pi}$.

Accordingly, I adopt a model in which the $\pi N$ channel is coupled to a $\pi^{2} N$ channel, where $\pi^{2}$ is treated as a quasiparticle with mass $m_{22} = 3m_{\pi}$. This model has been used previously in the analysis of $\pi N$ scattering data, with the $\pi^{2}$ mass adjusted for each partial wave \cite{Kiswandhi2004,Cutkovsky1979}.
%%%%%%%%%%%%%%%%%%%%%%%%%%%%%%%%%%%%%%%
The PWA data ($Re(T)$ and $Im(T)$ dependencies on $E_{lab}$ for the elastic $\pi N$ channel) allow one to calculate $S_{11}(q)=1+2\imath (Re(T)+\imath Im(T))$. From this, I obtained $\delta_{1}$ and $\cos(2\varepsilon)$.

For this model calculation, I assumed $\sin(2\varepsilon)\geq 0$ and $\delta_{2} \approx \delta_{1}/2$. According to my analysis, the function $\gamma$ was modeled as $\gamma=\alpha\sqrt{q|q_{2}|}$, where $\alpha\approx 0.07$ was extracted from $\sin(2\varepsilon)$ using Eqs.~(\ref{gamma_at_thres_left}) and (\ref{gamma_at_thres_right}). This value is only approximate because the data are not precise enough. The constant $c$ of Eq.~(\ref{gamma_at_q1=0}) is also unknown and was therefore chosen arbitrarily.

\begin{figure}[h]
	% Use the \centerline and \includegraphics commands to insert your figure file:
	\centerline{\includegraphics[width=0.5\textwidth]{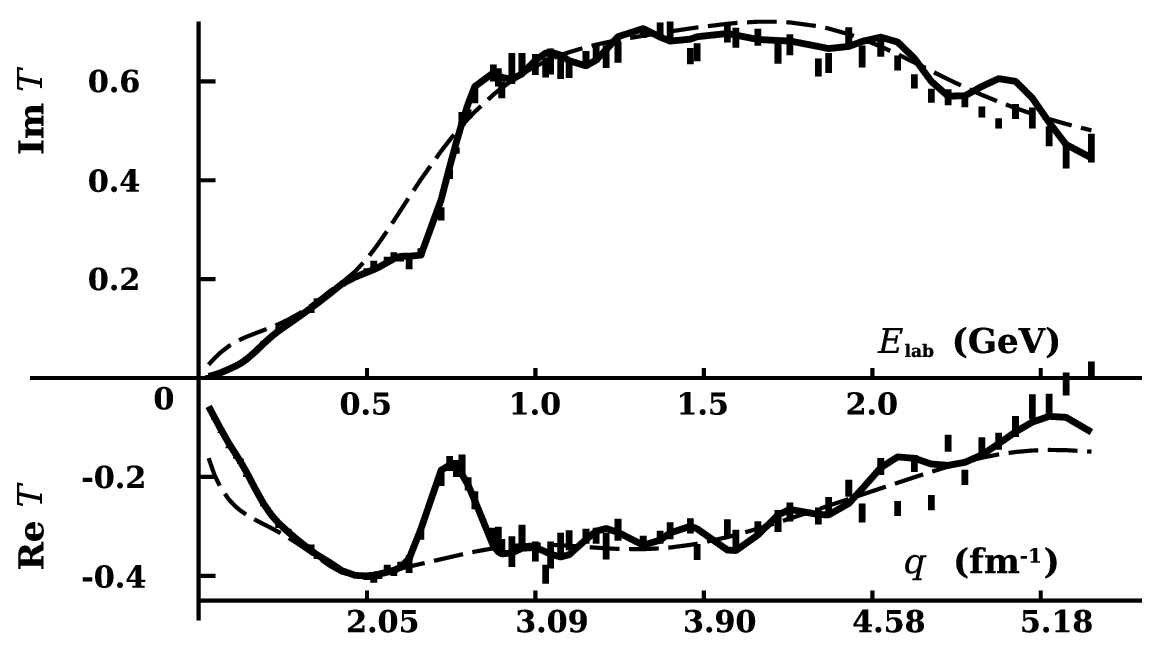}}
	%\centerline{\includegraphics[width=0.8\textwidth]{dwaves.eps}}
	% Use the \caption command to produce the figure caption and place it below the graph:
	\caption{\label{TS31old} 
				$S31\ \pi N$ $T$-matrix values (channel $\pi N \to \pi N$). Solid curves were calculated with inversion potentials (next figure) cut at $r=8$~fm, and dashed curves  were calculated with inversion potentials cut at $r=4$~fm;
		 data (rectangles) are from 
		\cite{TpiN_data}.
	}
\end{figure}

\begin{figure}[h]
	% Use the \centerline and \includegraphics commands to insert your figure file:
	\centerline{\includegraphics[width=0.5\textwidth]{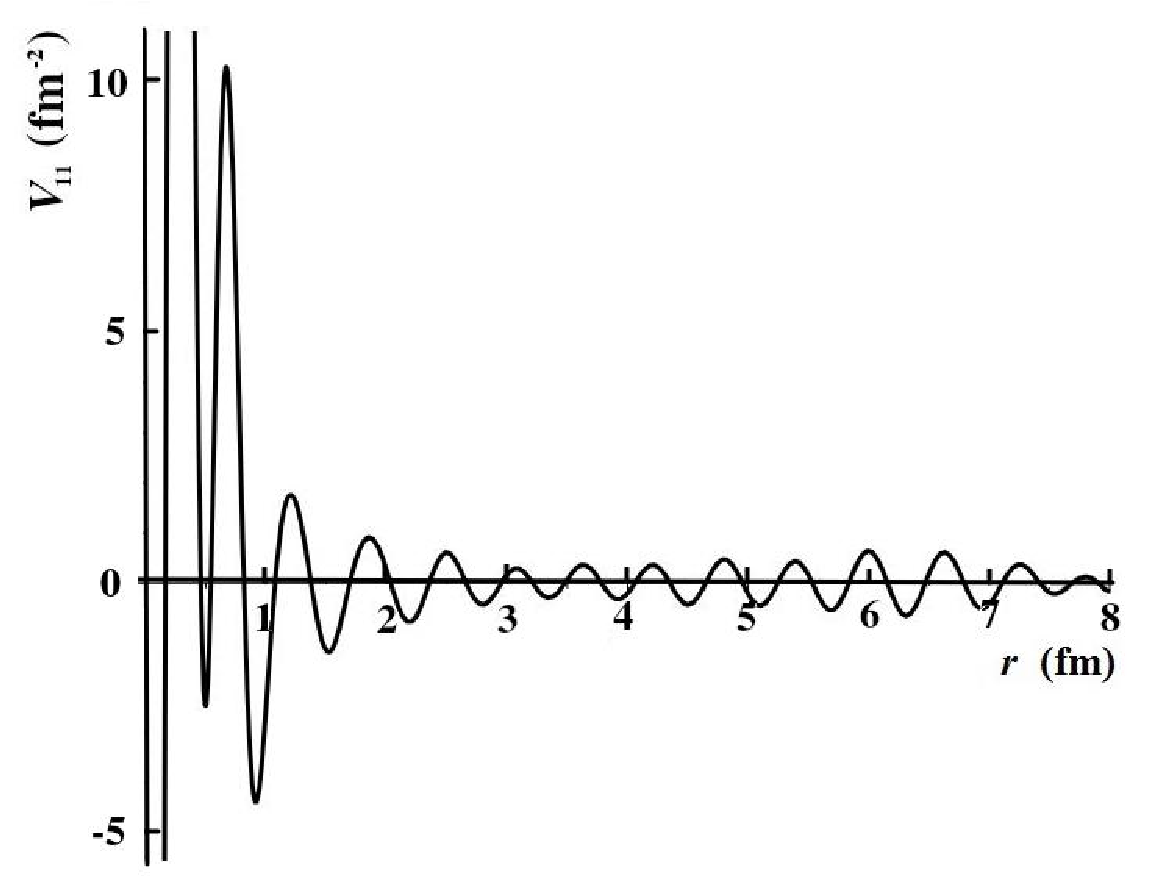}}
	\centerline{\includegraphics[width=0.5\textwidth]{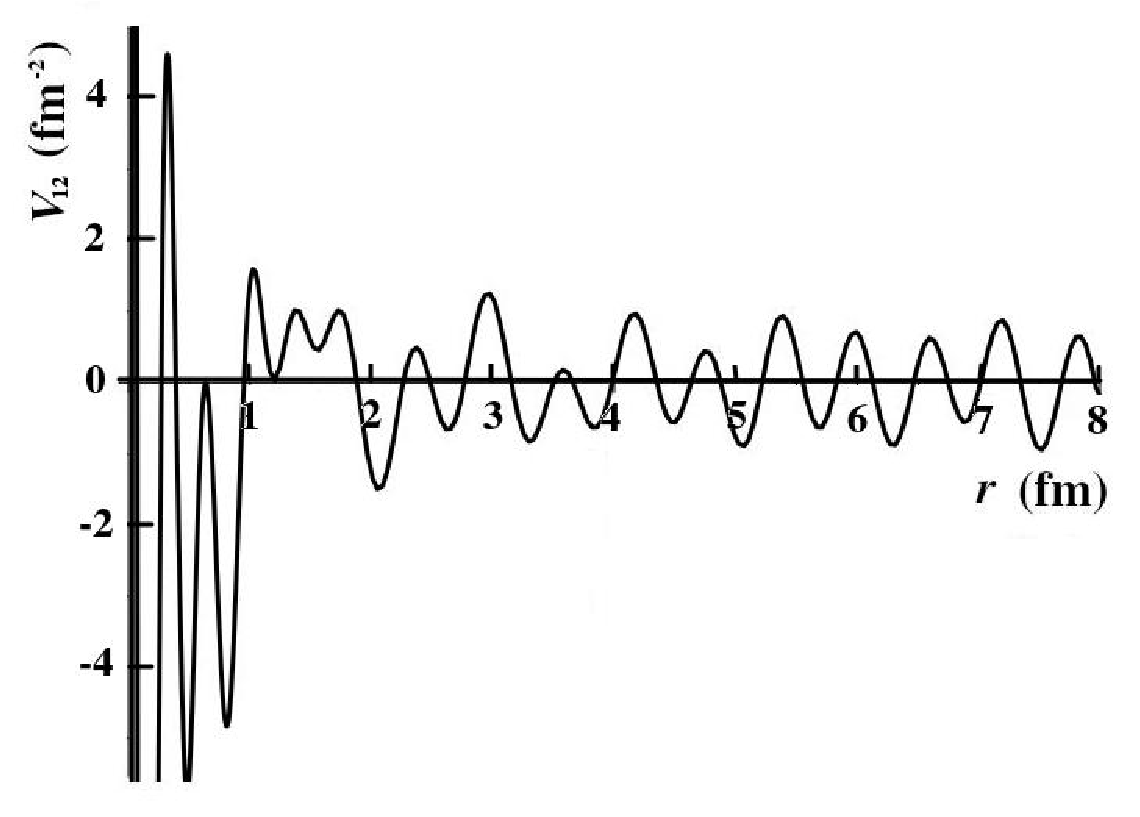}}
	\centerline{\includegraphics[width=0.5\textwidth]{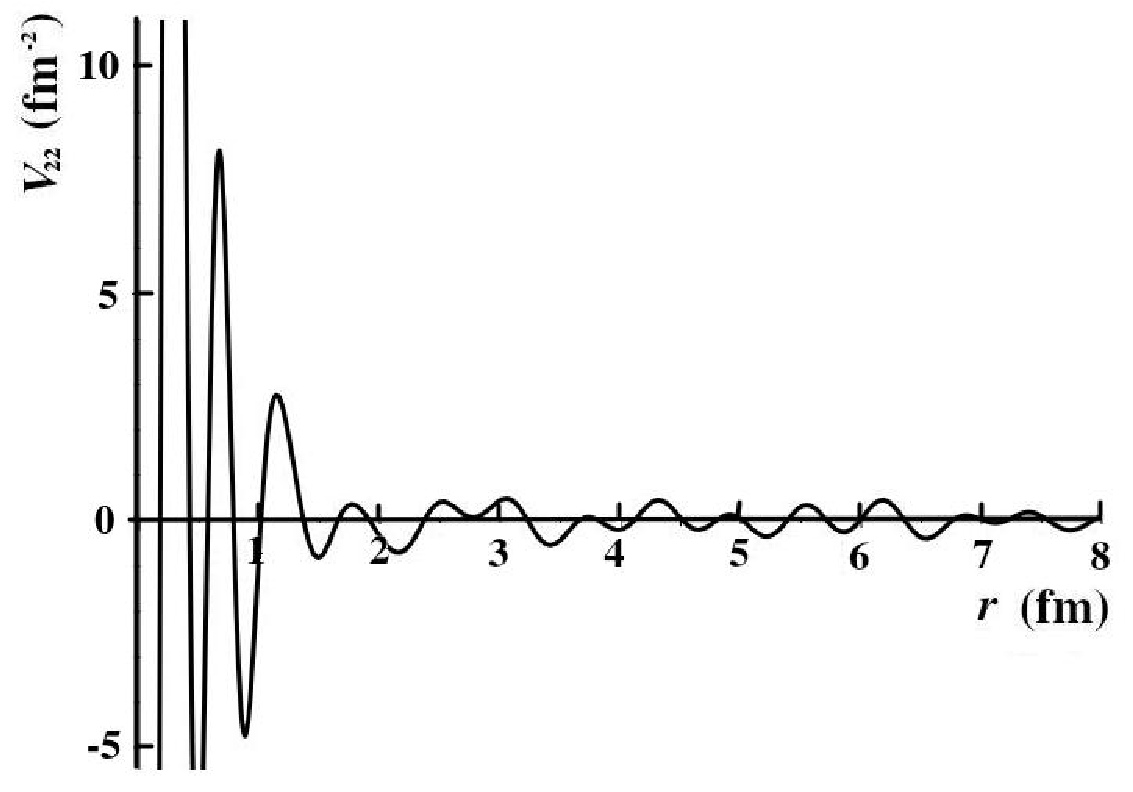}}
	%\centerline{\includegraphics[width=0.8\textwidth]{dwaves.eps}}
	% Use the \caption command to produce the figure caption and place it below the graph:
	\caption{\label{TS31pots} Inversion   $\pi N-\pi N^{3}$ potentials obtained from  data (rectangles) presented in Fig.~\ref{TS31old}.
	}
\end{figure}

Fig.~\ref{TS31old} shows the input PWA data alongside the results of the direct scattering calculation using the potentials obtained from the inverse problem solution.
 The resulting potentials, shown in Fig.~\ref{TS31pots}, exhibit significant oscillations and decay slowly toward $r=R\approx 13.95$~fm.

The direct-problem calculations presented in Fig.~\ref{TS31old} demonstrate, however, that a good description of the data—including the resonance region—is achieved when the potentials are cut off at $r=8$~fm. Cutting off the potentials at $r=4$~fm worsens the agreement at low energies and fails to reproduce the resonance region entirely.

%    	 The Marchenko equation is restricted to local potentials. If the locality condition
% is relaxed, an infinite number of solutions can yield the same $S$ matrix, binding energies,
% and asymptotic constants for bound states.

In summary, the developed method significantly extends my previous approach by providing a separable kernel for the Marchenko equation in the case of coupled channels with different thresholds. This enables both interpolation and extrapolation for any physically reasonable dependence of the corresponding $S$ matrix. In addition to generalizing the results of earlier work, new results have been obtained concerning the properties of the $S$ matrix for two-particle channels with different thresholds.

For the author, further advances in the description of $NN$ and $\pi N$ scattering are tied to the development of codes for problems involving three and more two-particle channels. Incorporating a more realistic description of three-particle channels into the developed formalism also remains a pressing task.
%&&&&&&&&&&&&&&&&&&&&&&&&&&&&&&&&&&&&&&&

The reconstructed $S_{31}$-wave $\pi N-\pi^{2} N$  potentials may be requested from the author in the Fortran code.

\section*{Acknowledgements}
During the preparation of this work, the author used DeepSeek AI for language editing and clarity improvements. After using this tool/service, the author reviewed and edited the content as needed and takes full responsibility for the content of the publication.
\section*{Data availability}
The original contributions presented in the study are included in
the article, further inquiries can be
directed to the author.


\end{document}